\documentclass[prb,aps,showpacs,twocolumn,superscriptaddress]{revtex4-1}
\usepackage{graphicx}
\usepackage{float}
\usepackage{epstopdf}
\usepackage{amsmath}
\usepackage{amssymb}
\usepackage{hyperref}
\usepackage{color}
\DeclareMathAlphabet{\mathpzc}{OT1}{pzc}{m}{it}
\DeclareFontFamily{OT1}{pzc}{}
\DeclareFontShape{OT1}{pzc}{m}{it}{ <-> s*[1.1] pzcmi7t }{}
\usepackage{contour}

\begin{document}

\author{Kallol Mondal}
\email{kallolsankarmondal@gmail.com}
\affiliation{School of Vocational Studies and Applied Sciences, Gautam Buddha University, Greater Noida, Uttar Pradesh, 201 312,  India}

\author{Sudin Ganguly}
\email{sudinganguly@gmail.com}
\affiliation{Department of Physics, Adamas University, Adamas Knowledge City, Barasat-Barrackpore Road, 24 Parganas North, Kolkata 700 126, India}

\author{Santanu K. Maiti}
\email{santanu.maiti@isical.ac.in}
\affiliation{Physics and Applied Mathematics Unit, Indian Statistical  Institute, 203 Barrackpore Trunk Road, Kolkata-700 108, India}

\title{Interplay of spin-orbit coupling and magnetic chirality: Multidirectional spin polarization in a helical antiferromagnet}
\begin{abstract}
We investigate spin-dependent transport in a non-collinear helical antiferromagnet in the presence of spin-orbit coupling within a tight-binding framework. Using the Landauer-B\"{u}ttiker formalism, we analyze the generation of spin polarization arising from the combined effects of the helical magnetic texture and spin-orbit interaction. We find that finite spin polarization can be generated simultaneously along the $\hat{x}$-, $\hat{y}$-, and $\hat{z}$-directions, in contrast to the predominantly single-axis spin polarization commonly observed in conventional spin-filtering systems. The multidirectional spin polarization originates from the symmetry breaking introduced by the non-collinear magnetic order. We further show that the spin polarization is modified significantly in the presence of spin-orbit coupling and it has strong dependence on the parameters associated with the helical system.  In particular, long-range hopping significantly enhances the spin-filtering effect, leading to large spin polarization components along all three spatial directions.

\end{abstract}

\maketitle
\section{\label{sec:intro}Introduction}
The efficient generation and control of spin-polarized currents are central objectives of spintronics because of their relevance to low-power electronics, nonvolatile memory technologies, and quantum information processing~\cite{spintronics1,spintronics2}. This pursuit has motivated intense research into magnetic materials capable of providing robust and versatile spin functionalities. Among them, antiferromagnets have recently emerged as promising candidates for spintronic applications~\cite{afm-rev1,afm-rev2,afm-rev3,afm-rev4}. Owing to their vanishing net magnetic moment, antiferromagnets are inherently insensitive to external magnetic-field perturbations and do not produce stray magnetic fields. In addition, their ultrafast spin dynamics enable device operation at frequencies far beyond those achievable in conventional ferromagnetic systems~\cite{fast1,fast2}. These advantages have established antiferromagnetic spintronics as an active and rapidly growing area of research.

The research in antiferromagnetic spintronics has primarily focused on conventional collinear magnetic configurations~\cite{kallol1,SuparnaPRB}, where magnetic moments are aligned antiparallel along a common axis. However, conventional antiferromagnets featuring only two magnetic atoms with antiparallel moments and no further local symmetry breaking between the sublattices exhibit spin-degenerate bands due to parity-time-reversal (${\mathcal {PT}}$) symmetry. Consequently, they lack a net spin polarization of the electronic band structure in momentum space~\cite{afm-rev3, afm-rev4}. Nevertheless, this symmetry can be lifted through external perturbations, enabling a variety of spin-transport phenomena in collinear antiferromagnets, including spin pumping, spin-current generation, and spin transport in engineered heterostructures~\cite{layer1,layer2,layer3,layer4,subtera,biasv}. These studies demonstrate that antiferromagnetic materials can serve as efficient active elements in spintronic devices when the underlying sublattice symmetry is appropriately broken. A more direct route to symmetry breaking is provided by non-collinear antiferromagnets, where the magnetic moments are not constrained to a common axis~\cite{kallol2,kallol3}. The reduced symmetry of such systems gives rise to a variety of emergent transport phenomena, including the anomalous Hall effect, spin Hall effect, inverse spin Hall effect, and anomalous Nernst effects~\cite{ahe1,ahe2,ahe3,ahe4,ahe5,she1,she2,she3,ane}. More recently, it has been demonstrated that non-collinear antiferromagnets can generate spin-polarized electrical currents as an intrinsic consequence of their symmetry-breaking magnetic texture~\cite{jakub1,jakub2}, analogous to the spin polarization observed in ferromagnetic systems in the presence of spin-orbit coupling.

Among the various non-collinear magnetic structures, helical antiferromagnets are particularly attractive because the continuous rotation of local magnetic moments generates a well-defined magnetic chirality. The reduced symmetry associated with such magnetic textures can significantly influence spin-dependent transport. These features make helical antiferromagnets a natural platform for investigating the combined effects of magnetic chirality and spin-orbit coupling on spin polarization. Insights into the role of helicity in spin transport have also emerged from studies of chiral molecular systems, where spin-selective transport can occur in the absence of magnetic ordering~\cite{Gohler}. In recent years, considerable attention has been devoted to chiral molecular systems, such as proteins and single-stranded DNA (ssDNA), which exhibit spin-selective transport properties commonly associated with the chiral-induced spin selectivity (CISS) effect~\cite{Guo-PRL,Guo-PNAS, Mishra-PNAS, Gutierrez,Pan, Guo-PRB}. Theoretical descriptions of the CISS effect often invoke spin-orbit coupling (SOC), arising from the interaction between the electron's spin and its orbital motion in a chiral environment~\cite{Guo-PNAS,Guo-PRB2}. These findings highlight the fundamental role of helicity and SOC in spin-selective transport and naturally motivate the investigation of their combined effects in a helical magnetic system.

In contrast to molecular helices, helical antiferromagnets inherently couple magnetic chirality with SOC, providing an ideal platform for spin-dependent transport. The coexistence of a non-collinear magnetic texture and SOC breaks the symmetries governing spin transport and can give rise to spin polarization beyond a single quantization axis. Consequently, helical antiferromagnets provide a promising platform for realizing and controlling multidirectional spin polarization. Despite these intriguing possibilities, the generation and manipulation of spin polarization along multiple spatial directions in non-collinear helical antiferromagnetic systems remain largely unexplored. Exploring these phenomena is therefore crucial for both a fundamental understanding of spin-dependent transport and the development of next-generation antiferromagnetic spintronic technologies.

In this work, we investigate spin-dependent transport in a non-collinear helical antiferromagnetic system with SOC within a tight-binding framework. The spin-dependent two-terminal transmission probabilities are computed using the well-known
Green's function formalism, based on Landauer-B\"{u}ttiker prescription~\cite{etms,qtat}.  \textit{We demonstrate that the interplay between helical magnetic order and SOC generates finite spin polarization, without invoking the dephasing mechanism}~\cite{Guo-PRB2}. Notably, spin polarization emerges simultaneously along the $\hat{x}$-, $\hat{y}$-, and $\hat{z}$-directions. We show that this multidirectional spin polarization originates from the combined breaking of spin-rotational and spatial-inversion symmetries by the non-collinear magnetic texture and SOC, and that its magnitude and orientation can be tuned through externally controllable helical parameters. These findings reveal a route toward coherent generation and manipulation of multidirectional spin polarization in chiral antiferromagnetic systems.

The rest of the work is organized as follows. In Sec.~\ref{sec:formalism}, we present our model system, having a non-collinear arrangement of the magnetic moments with a zero net moment. In this section, we also present a detailed theoretical description for the calculations of spin-resolved two-terminal transmission probability, currents, spin polarization coefficient for different helix configuration. All the results are critically investigated in Sec.~\ref{sec:result}. Finally, in Sec.~\ref{sec:conclusion}, we conclude our essential findings.

\section{\label{sec:formalism}Model and theoretical formulation}
\begin{figure}[t!]
\centering
\includegraphics[width=0.35\textwidth]{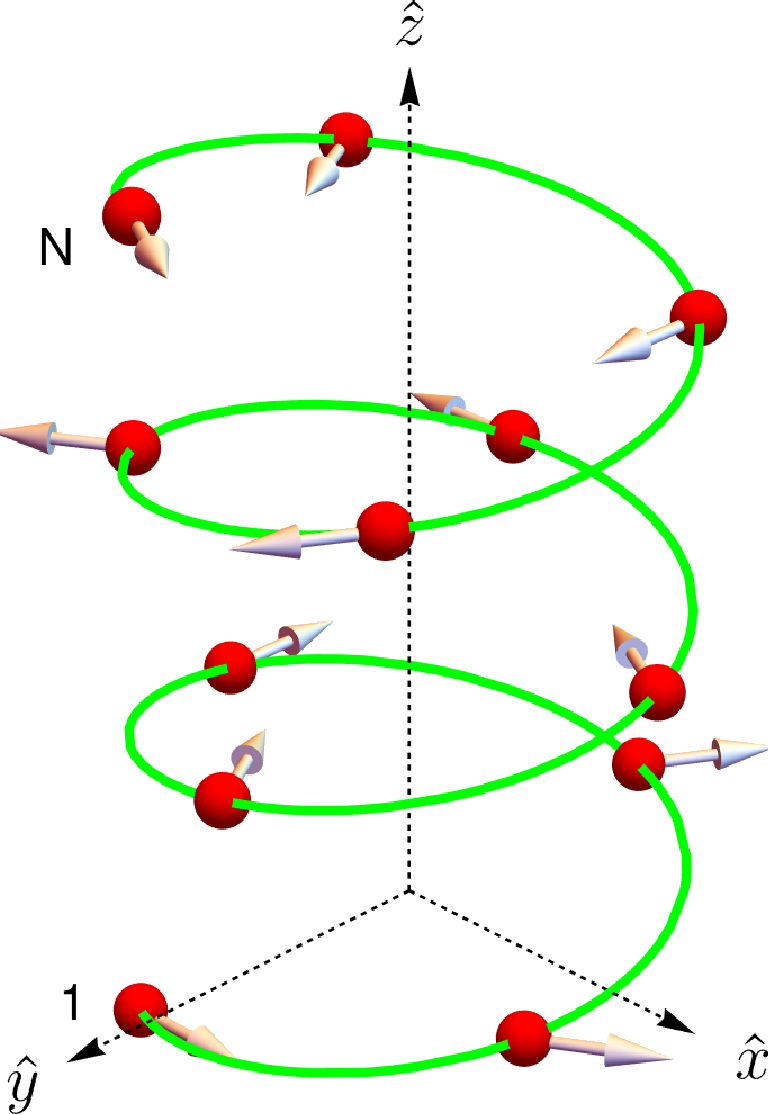} 
\caption{(Color online). Schematic representation of the non-collinear antiferromagnetic helix. Successive magnetic moments (shown with the 3D arrows associated with red balls) are progressively rotated with respect to one another, giving rise to a chiral magnetic texture with vanishing net magnetization. In the transport setup, the source and drain leads are attached to the end sites, denoted with 1 and N, respectively. The leads are omitted from the figure for clarity.}
\label{fig1:diagram}
\end{figure}

We consider a single-stranded non-collinear antiferromagnetic helix comprising $N$ magnetic sites, as shown schematically in Fig.~\ref{fig1:diagram}. The helix is right-handed and attached to two semi-infinite one-dimensional (1D)  non-magnetic leads, referred to as the source (S) and drain (D). The source lead is connected to site $1$, whereas the drain lead is connected to site $N$. 

Let us first discuss the geometry of the system. A helical structure is characterized by three geometric parameters: the radius $R$, the stacking distance $\Delta z$, and the twisting angle $\Delta \phi$~\cite{Guo-PNAS,Guo-PRL}. These parameters determine the spatial arrangement of the atomic sites and, consequently, the distances between them. Since the hopping amplitudes depend on the inter-site separation, the helical geometry plays a crucial role in determining the effective hopping range. Depending on the choice of $R$, $\Delta z$, and $\Delta \phi$, the system may exhibit either predominantly short-range hopping (SRH) or long-range hopping (LRH) characteristics. To investigate the influence of the hopping range on spin-dependent transport, we consider two representative helical geometries inspired by DNA and $\alpha$-helical proteins. For the SRH helix, we choose $R=7\,$\AA, $\Delta z=3.4\,$\AA,  and $\Delta\phi=\pi/5$. For the LRH helix, the corresponding parameters are $R=2.5\,$\AA, $\Delta z=1.5\,$\AA, and  $\Delta\phi=5\pi/9$. These parameter sets are employed solely to represent typical short-range and long-range hopping helices and are not intended to provide quantitative descriptions of actual DNA or $\alpha$-protein molecules~\cite{kallol1,kallolsci}. Further details of the helical geometry and the associated structural parameters regarding those biological systems can be found in Ref.~\cite{Guo-PNAS}.

We now turn to the magnetic structure of the system. The helix hosts a non-collinear antiferromagnetic spin texture in which the local magnetic moments rotate progressively from site to site while maintaining a vanishing net magnetization. Each site $n$ is associated with a local spin moment $\langle \mathbf{S}_n \rangle$, whose orientation is specified by the polar angle $\theta_n$ and the azimuthal angle $\varphi_n$. Without loss of generality, we choose the spin at the first site to be aligned along the $x$-axis. The spin configuration is thus defined as
\begin{subequations}
\begin{eqnarray}
\mathbf{S}_1 &=& \langle S \rangle \,\hat{\mathbf x}, \\
\mathbf{S}_n &=& \hat{R}\!\left(\hat{\mathbf z},\Phi_n\right)\cdot \mathbf{S}_1,
\end{eqnarray}
\label{eq:spin-mag}%
\end{subequations}
where $\hat{R}(\hat{\mathbf z},\Phi_n)$ denotes the rotation operator corresponding to a rotation by an angle  $
\Phi_n=\frac{2  \pi (n-1)}{N}, $  about the helix axis $\hat{\mathbf z}$, and $\langle S\rangle$ is the magnitude of the local spin moment. With this choice, the magnetic moments rotate smoothly along the helix, resulting in a non-collinear antiferromagnetic configuration with zero net magnetization i.e.
\begin{equation}
\sum_{n=1}^N \mathbf{S}_n =0.
\end{equation}
\subsection{Model Hamiltonian}

The total Hamiltonian of the system consists of three parts: the central non-collinear antiferromagnetic helix~\cite{Guo-PNAS,Guo-PRB}, the source and drain leads, and the coupling between the helix and the leads. It is expressed as
\begin{equation}
\mathcal{H} = \mathcal{H}_{\text{helix}} + \mathcal{H}_{\text{leads}} + \mathcal{H}_{\text{coupling}}.
\end{equation}

\vspace{0.2cm}
\noindent \textbf{Helix Hamiltonian:} In our model, the non-collinear antiferromagnetic helix is described within a tight-binding framework that incorporates  spin-dependent scattering (SDS), electron hopping, and the spin-orbit coupling respectively. The Hamiltonian of the considered system is given by
\begin{align}
&\mathcal{H}_{\rm helix} = \sum_{n=1}^{N} \mathbf{c}_n^\dagger \left( \boldsymbol{\epsilon}_n - \contour[1]{black}{$\mathpzc{h}$}_n\cdot\boldsymbol{\sigma} \right) \mathbf{c}_n \nonumber\\
&+  \sum_{n=1}^{N-1} \sum_{m=1}^{N-n} \left( \mathbf{c}_n^\dagger \mathbf{t}_{n}\mathbf{c}_{n+m} +{\rm h.c.}\right)  \nonumber\\
 &+  \sum_{n=1}^{N-1} \sum_{m=1}^{N-n}  2i s_m \cos(\phi^-_{nm}) \left( \mathbf{c}_n^\dagger \boldsymbol{\sigma}_{nm}\mathbf{c}_{n+m} +{\rm h.c.}\right) ,
\label{eq:Hh}
\end{align}
where $ \mathbf{c}_n^\dagger = \left( c_{n\uparrow}^{\dagger}, c_{n\downarrow}^{\dagger} \right) $ and $ \mathbf{c}_n = \left(
c_{n\uparrow}, c_{n\downarrow} \right)^T $ are the spinor creation and annihilation operators at site $n$, respectively, and
$\boldsymbol{\sigma}=(\sigma_x,\sigma_y,\sigma_z)$ is the vector of Pauli matrices. The matrices $\boldsymbol{\epsilon}_n$ and $\mathbf{t}_n$ are defined as
\begin{equation}
\boldsymbol{\epsilon}_n=
\begin{pmatrix}
\epsilon_n & 0\\
0 & \epsilon_n
\end{pmatrix},
\qquad
\mathbf{t}_n=
\begin{pmatrix}
t_n & 0\\
0 & t_n
\end{pmatrix},
\end{equation}
where $\epsilon_n$ denotes the on-site energy in the absence of any spin-dependent scatterings and $t_n$ represents the hopping amplitude between sites $n$ and $(n+m)$.  

The first term in the helix Hamiltonian $\mathcal{H}_\text{helix}$ describes the spin-dependent scattering of the incoming electrons. Here, an incoming electron interacts with the local magnetic moments through an exchange coupling $J$. This interaction is incorporated via spin-dependent scattering (SDS) field~\cite{YHsu}, defined as
\begin{equation}
\contour[1]{black}{$\mathpzc{h}$}_n= J \langle \mathbf{S}_n \rangle,
\end{equation}
where $\langle \mathbf{S}_n \rangle$ denotes the local magnetic moment at site $n$. For simplicity, the magnitude of the SDS field is assumed to be uniform throughout the helix, i.e., $|\contour[1]{black}{$\mathpzc{h}$}_n| = \mathpzc{h}$ $\forall n$, while its direction follows the underlying non-collinear magnetic texture. Furthermore, it is important to note that direct interactions between neighboring localized magnetic moments are neglected in the present model. This is because in typical transition-metal helimagnets, stabilizing localized interactions (e.g., Heisenberg exchange and Dzyaloshinskii-Moriya interactions) are typically in the meV range~\cite{heide, zimm}. In contrast, the relevant electronic energy scales operate in the eV regime. These include the hopping integral ($t$) and the local exchange field ($\mathpzc{h}$). Due to this large energy difference, the moment-moment interactions do not significantly alter the electronic band structure and the transport properties. The effect of SDS enters through the term $\contour[1]{black}{$\mathpzc{h}$}_n\cdot\boldsymbol{\sigma}$, where $\contour[1]{black}{$\mathpzc{h}$}_n$ is the SDS field associated with site $n$. Consequently, the effective on-site potential becomes
$
\boldsymbol{\epsilon}_n-\contour[1]{black}{$\mathpzc{h}$}_n\cdot\boldsymbol{\sigma}
$,
which couples the electron spin to the underlying non-collinear magnetic configuration. 

The second term in the helix Hamiltonian $\mathcal{H}_\text{helix}$ describes the hopping between the sites. The nature of helical geometry makes the hopping integral a bit tricky unlike usual nearest-neighbor hopping case. The expression for the hopping integral $t_n$ is given by
\begin{equation}
t_n = t_1 \exp\left[-\frac{(l_n-l_1)}{l_c}\right]
\end{equation}
where $t_1$ and  $l_1$ are the nearest-neighbor hopping amplitude,  and the distance among the nearest-neighbor sites respectively.
$l_n$ is the spatial separation between the sites $n$ and $n+m$. The expression of $l_n$ is given by
\begin{equation}
l_n = \sqrt{\left( 2R \sin\left(n \Delta \phi/2\right)\right)^2+ \left( n \Delta z \right)^2},
\end{equation}
where $\Delta z$ and $\Delta \phi$ are the stacking distance and twisting angle respectively, as discused earlier. $l_ c$ is the decay constant, and following the Ref.~\cite{Guo-PNAS}, it is taken to be $0.9~\text{\AA}$ for both the  SRH and LRH cases.  
 
The final term in the helix Hamiltonian, $\mathcal{H}_{\rm helix}$, accounts for SOC arising from the helical geometry. The SOC strength is assumed to depend on the inter-site separation and is modeled in a form analogous to the hopping integral,
\begin{equation}
s_m=s_1\exp\left[-\frac{(l_m-l_1)}{l_c}\right],
\label{eq:soc}
\end{equation}
where $s_1$ denotes the nearest-neighbor SOC strength, $l_1$ is the nearest-neighbor distance, and $l_c$ is the decay length introduced previously and $i$ is $\sqrt{-1}$. The spin-dependent coupling matrix $\sigma_{nm}$ is given by
\begin{equation}
\sigma_{nm}=\left(\sigma_x \sin\phi_{nm}^{+}-\sigma_y \cos\phi_{nm}^{+}\right)\sin\theta_m+\sigma_z \cos\theta_m,
\label{eq:gamma}
\end{equation}
where 
\begin{equation}
\phi_{nm}^{\pm} =\frac{\phi_{n+m}\pm\phi_n}{2},\qquad \phi_n=n\Delta\phi.
\end{equation}

Throughout this work, the SOC strength is measured in units of the nearest-neighbor hopping amplitude $t_1$. Accordingly, we define a parameter $t_{\rm so}\equiv s_1/t_1$, which is used to characterize the relative strength of spin-orbit coupling, and its values are specified where appropriate.

\vskip 0.1 in
\noindent\textbf{Lead Hamiltonian:}
The source and drain leads are modeled as semi-infinite 1D non-magnetic leads described by 
\begin{align}
&\mathcal{H}_{\rm S} = \sum_{m<1} \mathbf{a}_m^\dagger\boldsymbol{\epsilon}_0\mathbf{a}_m + \sum_{m<1}\left(\mathbf{a}_m^\dagger\mathbf{t}_0\mathbf{a}_{m-1}+{\rm h.c.}\right),\\
&\mathcal{H}_{\rm D} = \sum_{m> N}\mathbf{b}_m^\dagger\boldsymbol{\epsilon}_0\mathbf{b}_m + \sum_{m>N}\left(\mathbf{b}_m^\dagger\mathbf{t}_0\mathbf{b}_{m+1}+{\rm h.c.}\right),\\
&\mathcal{H}_{\rm leads} = \mathcal{H}_{\rm S}+\mathcal{H}_{\rm D}.
\end{align}

Here, $\mathbf{a}_m$ and $\mathbf{b}_m$ are the spinor annihilation operators associated with the source and drain electrodes, respectively, and are defined analogously to the operator $\boldsymbol{c}_n$ introduced for the magnetic helix. The matrices $\boldsymbol{\epsilon}_0$ and $\mathbf{t}_0$ denote the spin-independent on-site potential and nearest-neighbor hopping matrices of the leads, respectively. For simplicity, identical parameters are assumed for both the leads.
 
 \vspace{0.5cm}

\noindent\textbf{Coupling Hamiltonian:}
The coupling between the magnetic helix and the leads is described by
\begin{equation}
\mathcal{H}_{\rm coupling} =\mathbf{a}_{0}^{\dagger}\boldsymbol{\tau}_{\rm S}\mathbf{c}_{1}+\mathbf{c}_{N}^{\dagger}\boldsymbol{\tau}_{\rm D}\mathbf{b}_{N+1}+
{\rm h.c.},
\end{equation}
where $\boldsymbol{\tau}_{\rm S}$ and $\boldsymbol{\tau}_{\rm D}$ denote the coupling matrices between the source lead and the first site of the helix, and between the drain lead and the last site of the helix, respectively. Similar to the lead hopping matrix $\mathbf{t}_0$, both $\boldsymbol{\tau}_{\rm S}$ and $\boldsymbol{\tau}_{\rm D}$ are taken to be spin-independent $2\times2$ diagonal matrices.

\subsection{Spin-dependent transmission probabilities}

To investigate spin-dependent transport through the non-collinear antiferromagnetic helix, we employ the nonequilibrium Green's function (NEGF) formalism~\cite{etms,qtat,Landauer, Fisher}. The transport properties of the system are characterized by the spin-resolved transmission probabilities, which quantify the likelihood of an electron entering the magnetic helix from the source lead with spin $\sigma$ and exiting through the drain lead with spin $\sigma^\prime$. In the conventional $z$-axis spin basis, the spin-dependent transmission probability~\cite{etms,qtat,Landauer} is given by
\begin{equation}
T_{\sigma\sigma'}  = {\rm Tr} \left[ \Gamma_{\rm S}^{\sigma} \mathcal{G}^{r} \Gamma_{\rm D}^{\sigma^\prime} \mathcal{G}^{a}\right],\quad\quad(\sigma,\sigma^\prime = \uparrow,\downarrow),
\label{eq:trans}
\end{equation}
where $\Gamma_{\rm S}^{\sigma}$ and $\Gamma_{\rm D}^{\sigma^\prime}$ are the coupling matrices associated with the source and drain leads for the $\sigma$ and $\sigma^\prime$ channels, respectively. These matrices are related to the lead self-energies through 
\begin{equation}
\Gamma_{\alpha}^{\sigma} =-2\,{\rm Im}\left(\Sigma_{\alpha}^{\sigma}\right),\qquad \alpha={\rm S,D}.
\end{equation}

The matrices $\mathcal{G}^{r}$ and $\mathcal{G}^{a}$ denote the retarded and advanced Green's functions of the helix, satisfying $\mathcal{G}^{a}=(\mathcal{G}^{r})^{\dagger}$. The retarded Green's function is expressed as
\begin{equation}
\mathcal{G}^{r} = \left[ E\mathbf{I} - \mathcal{H}_{\rm helix} - \Sigma_{\rm S} - \Sigma_{\rm D} \right]^{-1},
\label{eq:green}
\end{equation}
where $E$ is the energy of the incident electron and $\mathbf{I}$ is the identity matrix of appropriate dimension.

The quantities $T_{\uparrow\uparrow}$ and $T_{\downarrow\downarrow}$ correspond to spin-conserving transmission probabilities, whereas $T_{\uparrow\downarrow}$ and $T_{\downarrow\uparrow}$ represent spin-flip transmission induced by the combined action of the non-collinear magnetic texture and SOC. The total transmission probabilities associated with spin-up and spin-down electrons collected at the drain are defined as
\begin{equation}
T_{\uparrow}=T_{\uparrow\uparrow}+T_{\downarrow\uparrow},\qquad T_{\downarrow}=T_{\downarrow\downarrow}+ T_{\uparrow\downarrow}.
\label{eq:totaltrans}
\end{equation}

\subsection{Spin-dependent transport currents and spin polarization}

In the presence of a non-collinear magnetic texture and SOC, the transmitted electron spin is generally not restricted to a single quantization axis. Consequently, a complete characterization of spin transport requires the evaluation of the spin components along all three spatial directions ($\alpha=x,y,z$). To gain deeper insight into these directional transport properties, we evaluate the spin-resolved transmission probabilities along the three Cartesian axes. The spin-up ($\uparrow$) and spin-down ($\downarrow$) transmission probabilities with respect to an arbitrary spin quantization axis $\alpha$ are obtained by projecting the full transmission matrix onto the corresponding spin eigenstates as~\cite{niko2005, dalum}
\begin{equation}
T_{\alpha}^{\uparrow(\downarrow)}={\rm Tr}\left[\mathcal{P}_{\alpha}^{\uparrow(\downarrow)}\Gamma_{\rm S}\mathcal{G}^{r}\Gamma_{\rm D}\mathcal{G}^{a}\right],
\label{eq:Talphapm}
\end{equation}
where
\begin{equation}
\mathcal{P}_{\alpha}^{\uparrow(\downarrow)}=I_N \otimes\frac{1}{2}\left(I_2 \pm \sigma_\alpha\right)
\end{equation}
represents the projection operators onto the spin-up ($+$ sign) and spin-down ($-$ sign) states along the $\alpha$-direction.  Here, $\sigma_\alpha$ denotes the corresponding Pauli matrix, while $I_N$ and $I_2$ denote the identity matrices in the site and spin subspaces, respectively. The total transmission probability is recovered from $T = T_{\alpha}^{\uparrow} + T_{\alpha}^{\downarrow}$, while the net spin-polarized transmission along the $\alpha$-direction is given by~\cite{po-chang, riv}
\begin{equation} 
T_{\alpha}^{s} = T_{\alpha}^{\uparrow} - T_{\alpha}^{\downarrow} = {\rm Tr}\left[\sigma_\alpha\Gamma_{\rm S}\mathcal{G}^{r}\Gamma_{\rm D}\mathcal{G}^{a}\right].
\label{tspin}
\end{equation}

Once these directional transmission probabilities are established, the corresponding spin-dependent transport currents can be evaluated within the Landauer-B\"uttiker framework. At finite temperatures, the current associated with the spin state $\uparrow(\downarrow)$ along the $\alpha$-axis is given by
\begin{equation}
I^{\uparrow(\downarrow)}_\alpha(V) = \frac{e}{h} \int_{-\infty}^{\infty} T^{\uparrow(\downarrow)}_\alpha(E) \left[ f_{\rm S}(E)-f_{\rm D}(E) \right] \, dE ,
\label{eq:finiteTcurrent}
\end{equation}
where $f_{\rm S}(E)$ and $f_{\rm D}(E)$ are the Fermi-Dirac distribution functions of the source and drain leads, respectively:
\begin{equation}
f_{\nu}(E) =\frac{1}{1+\exp\!\left[(E-\mu_{\nu})/k_{\rm B}T\right]},\qquad\nu={\rm S,D}.
\end{equation}
Here, $k_{\rm B}$ is the Boltzmann constant, $T$ is the temperature, and the electrochemical potentials of the leads are taken as $\mu_{\rm S}=E_F+eV/2$ and $\mu_{\rm D}=E_F-eV/2$, with $E_F$ denoting the equilibrium Fermi energy and $V$ representing the applied bias voltage.

The total charge current $I$ and the $\alpha$-component of the spin current $I_\alpha^s$ are subsequently determined from these spin-resolved currents via $I = I_\alpha^\uparrow + I_\alpha^\downarrow$ and $I_\alpha^s = I_\alpha^\uparrow - I_\alpha^\downarrow$. In the zero-temperature limit, the Fermi functions reduce to step functions, and the expression for the directional spin current simplifies to
\begin{equation}
I_\alpha^s = \frac{e}{h} \int_{E_F-\frac{eV}{2}}^{E_F+\frac{eV}{2}} T_\alpha^s(E)\,dE.
\label{spin-curr}
\end{equation}

We define the directional spin polarization coefficient $P_\alpha$ as~\cite{rai-prb-2012,mp}
\begin{equation}
P_\alpha = \frac{I_\alpha^s}{I} \times 100\%.
\label{eq:polarization}
\end{equation}
The polarization coefficient satisfies $-100\% \le P_\alpha \le 100\%$, where $P_\alpha=0$ corresponds to the absence of spin polarization, while $|P_\alpha|=100\%$ indicates complete spin polarization along the $\alpha$-direction. 
\section{\label{sec:result}Numerical results and discussion }
\begin{figure*}[ht]
\includegraphics[width=0.33\textwidth]{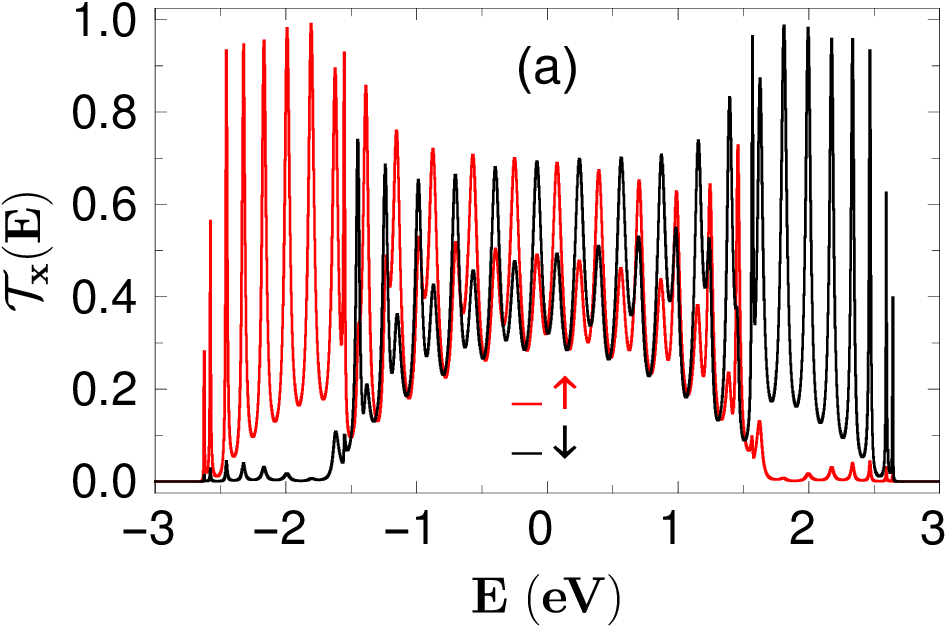}\hfill
\includegraphics[width=0.33\textwidth]{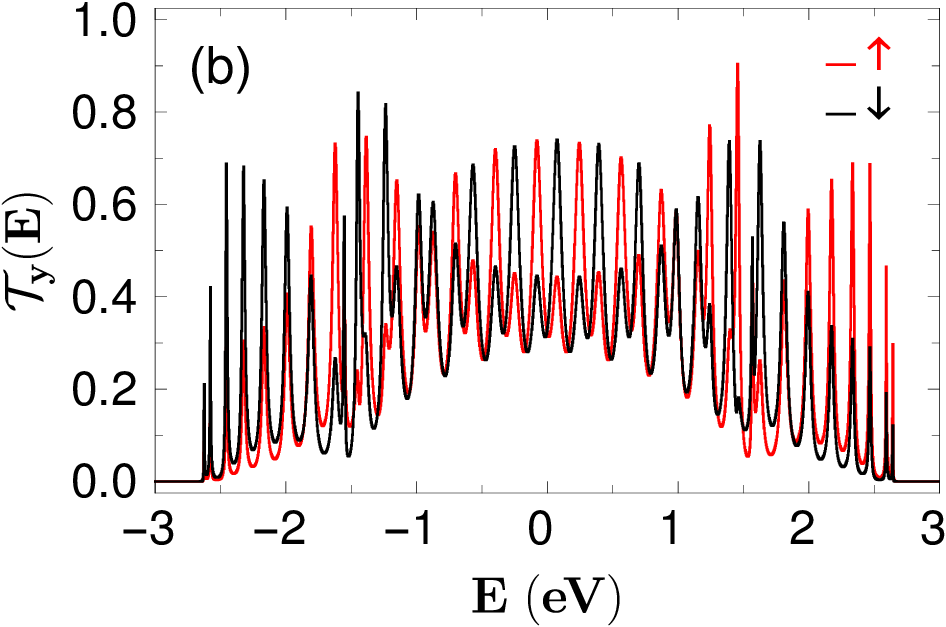}\hfill
\includegraphics[width=0.33\textwidth]{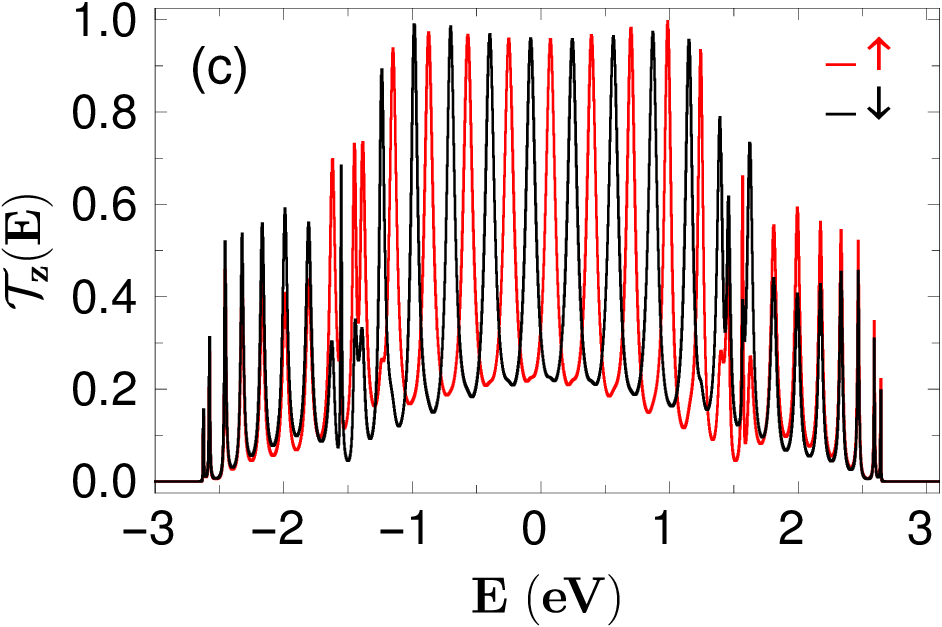}
\caption{(Color online). Energy dependence of the spin-resolved transmission probabilities ${\mathcal{T}}_\alpha^{\uparrow(\downarrow)}$ $(\alpha=x,y,z)$ for the short-range hopping (SRH) helix with $N=20$, $t_{\rm so}=0.25$, and $\mathpzc{h}=0.5$. Panels (a), (b), and (c) correspond to the $x$, $y$, and $z$-directions, respectively. In each panel, the red (black) curves represent the spin-up (spin-down) transmission probabilities.}
\label{Fig2}
\end{figure*}
\begin{figure*}[ht]
\includegraphics[width=0.32\textwidth]{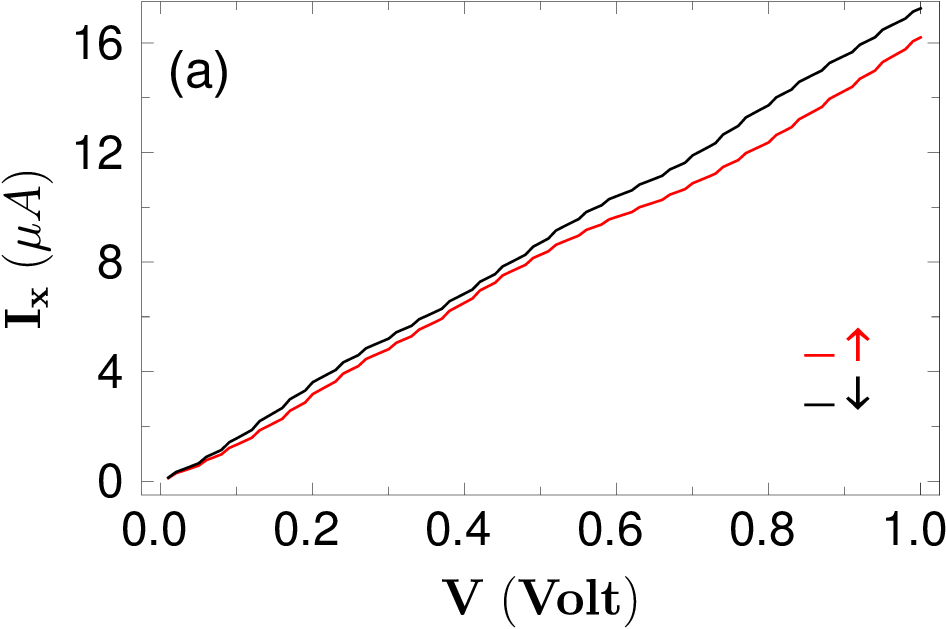}\hfill
\includegraphics[width=0.32\textwidth]{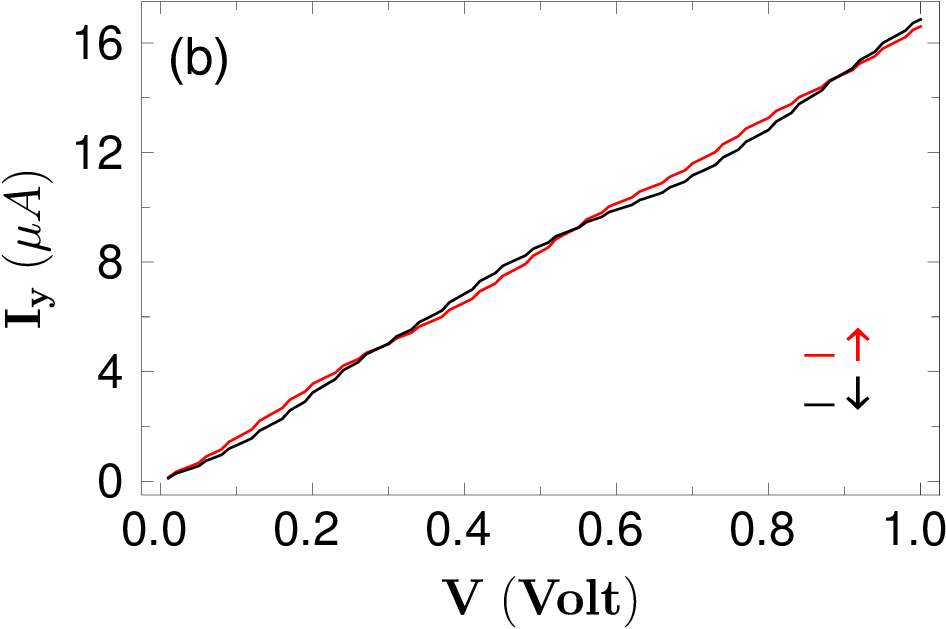}\hfill
\includegraphics[width=0.32\textwidth]{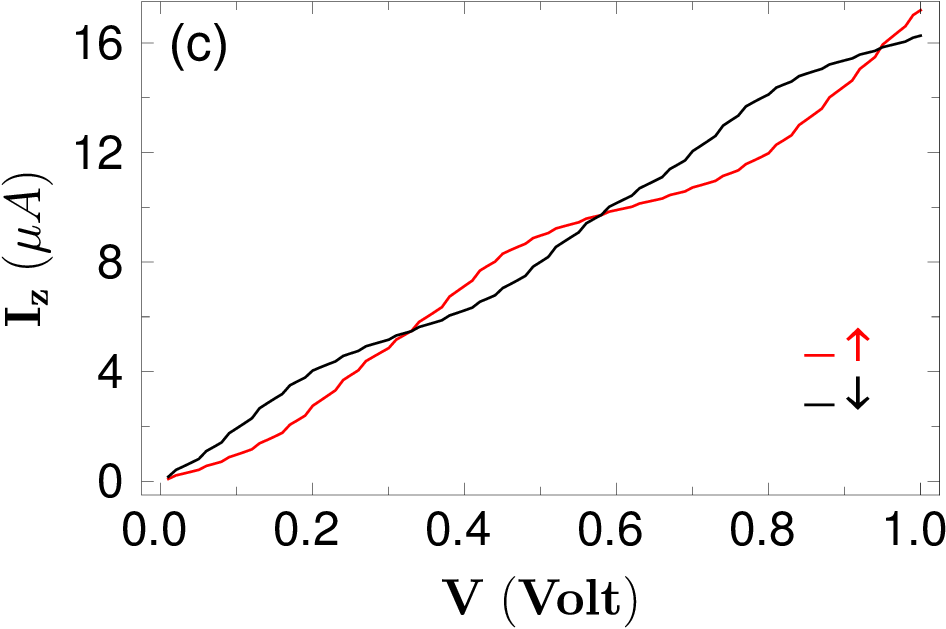}
\caption{(Color online). Spin-resolved current components $I_\alpha^{\uparrow(\downarrow)}$ $(\alpha=x,y,z)$ as a function of the applied bias voltage for the SRH configuration, shown in panels (a), (b), and (c), respectively. The red and black curves correspond to the up- and down-spin currents. All data are computed for a fixed Fermi energy $E_F=0.5\,$eV. The system parameters are the same as Fig.~\ref{Fig2}.}
\label{Fig3}
\end{figure*}
\begin{figure*}[ht]
\includegraphics[width=0.32\textwidth]{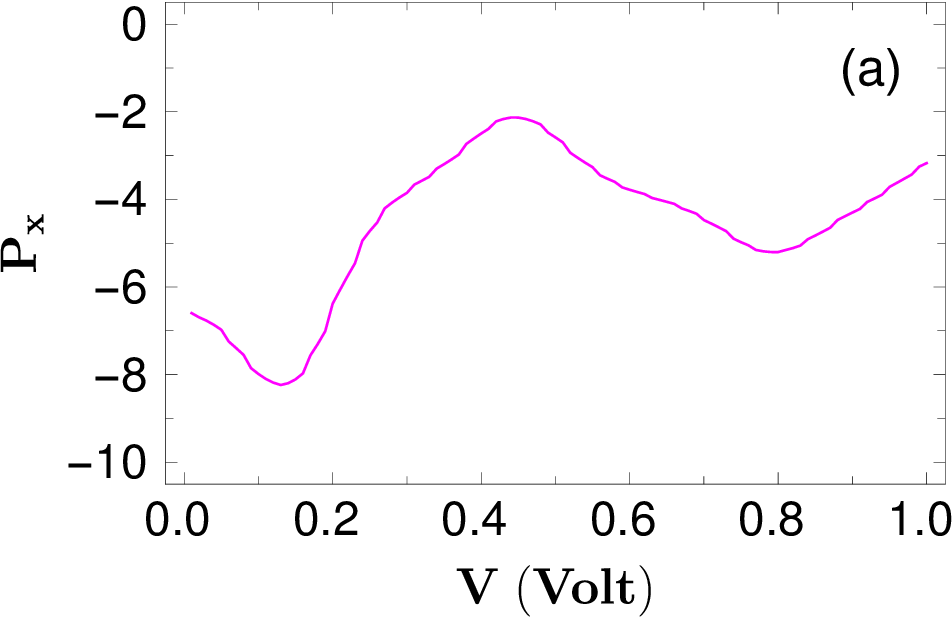}\hfill
\includegraphics[width=0.32\textwidth]{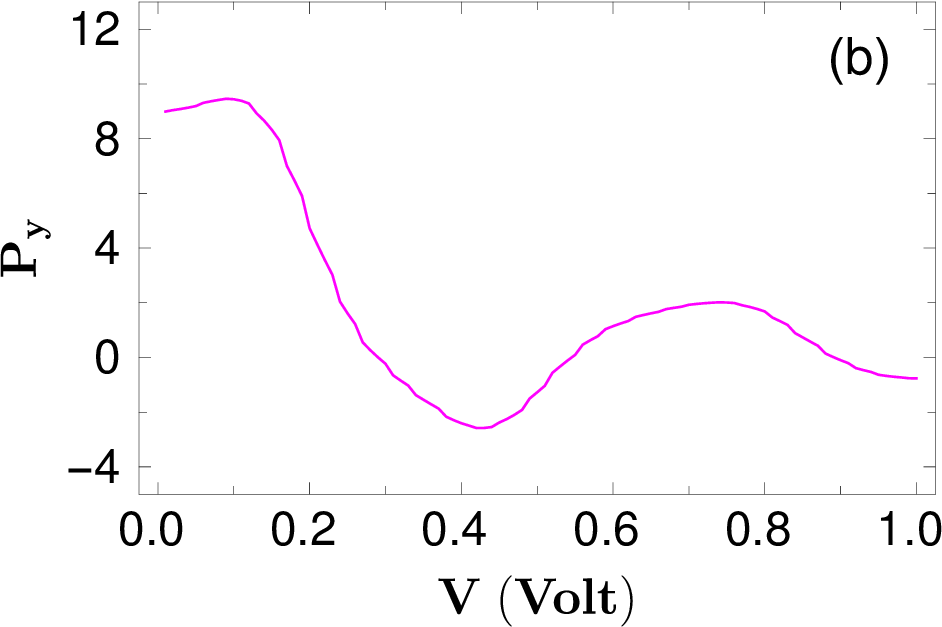}\hfill
\includegraphics[width=0.32\textwidth]{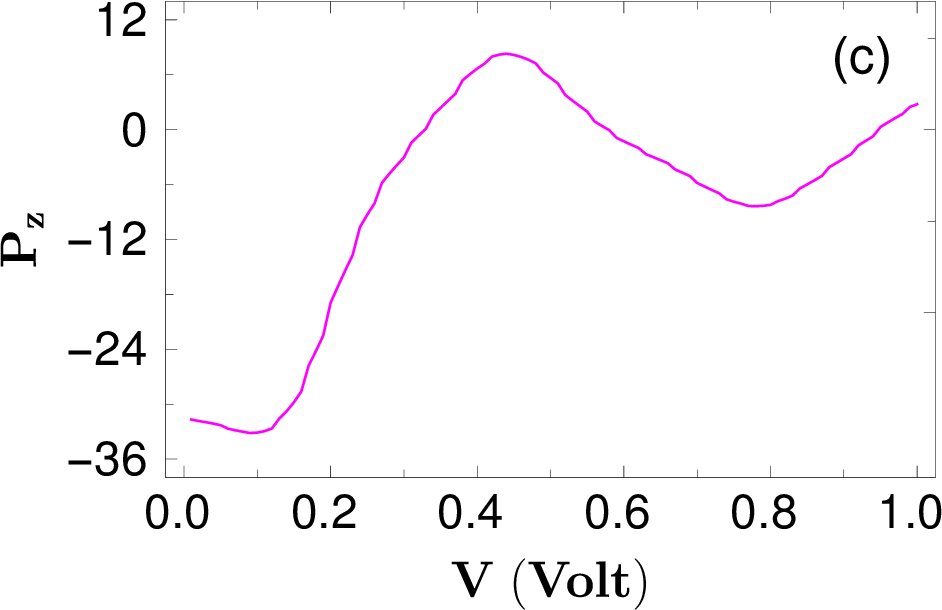}
\caption{(Color online). Bias-dependent spin polarization $P_\alpha$ $(\alpha=x,y,z)$ in the SRH configuration. (a)-(c) Show the spin polarization coefficients $P_x$, $P_y$, and $P_z$ as functions of the applied bias voltage $V$. The system parameters are the same as Fig.~\ref{Fig2}.}
\label{Fig4}
\end{figure*}

Before presenting the numerical results, we briefly summarize the parameter values used throughout this work. Unless otherwise stated, all energies are measured in units of eV. The on-site energies of the helical system and the source and drain leads are set to zero, i.e., $\epsilon_n=\epsilon_0=0$. The nearest-neighbor hopping integral in the leads is taken as $t_0=2.5$ eV, while the coupling strengths between the helix and the source (drain) lead is chosen as $\tau_{\rm S}=\tau_{\rm D}=0.8$ eV.  The above choice of hopping and coupling parameters allows us to work in the wide-band limit. It is important to emphasize, however, that the qualitative physical picture remains unchanged for other reasonable choices of the model parameters, as confirmed by our extensive numerical calculations. The remaining parameters, including the SOC strength and spin-dependent scattering strength, are specified at the appropriate places in the text. We further assume that the level broadening induced by the coupling between the helix and the leads is much larger than the thermal broadening $k_B T$. Consequently, the transport calculations presented below are performed within the zero-temperature approximation. Unless explicitly specified, all results are evaluated at absolute zero temperature. The influence of finite temperature on the spin polarization will be discussed separately for completeness.


\subsubsection{Short-range hopping scenario }
We begin our discussion of the numerical results by analyzing the spin-dependent transmission probabilities for the SRH helix. Because the spin polarization and the associated spin currents are directly determined by the underlying transmission characteristics, examining the spin-dependent transmission spectra provides essential insight into the microscopic origin of the spin-selective transport. Figure~\ref{Fig2} displays the energy dependence of the spin-resolved transmission probabilities along the three Cartesian directions. Panels (a)-(c) correspond to the result of $x$-, $y$-, and $z$-axes, respectively, where the red (black) curves denote the spin-up (spin-down) transmission probabilities. The calculations are performed for an SRH configuration with $N=20$, $t_{\rm so}=0.25$, and $\mathpzc{h}=0.5$. Crucially, the transmission probabilities associated with the spin-up and spin-down channels are unequal over a wide range of energies across all three panels, indicating robust spin-selective transport through the helix.

In Fig.~\ref{Fig2}(a), a pronounced spin splitting appears near the band edges along the $x$-direction, where one spin channel exhibits a substantially larger transmission than the other. In contrast, the transmission spectra along the $y$-direction do not show such strong splitting near the band edges, instead, the difference between the two spin channels is distributed across the central part of the transmission spectrum, as shown in Fig.~\ref{Fig2}(b). A distinct behavior is observed along the $z$-direction (Fig.~\ref{Fig2}(c)), where a clear separation between the spin-up and spin-down transmission probabilities persists over a broad energy range. These spectra confirm that a finite spin polarization is simultaneously generated along all three directions.

It is worth noting that the energy windows hosting the strongest spin-channel asymmetry vary significantly among the three components. While the splitting is highly concentrated near the band edges for the $x$-component, it remains active over a much broader, central energy range for the $y$- and $z$-components, suggesting that the resulting spin transport properties will exhibit distinct sensitivities to changes in the Fermi level. Ultimately, the simultaneous appearance of spin-channel asymmetry along all three axes reflects the combined influence of the non-collinear magnetic texture and SOC. The helical antiferromagnetic ordering breaks the local spin-sublattice symmetry, while the SOC mixes different spin components. Consequently, we get finite spin polarization along the $x$, $y$, and $z$ directions. Furthermore, the spin-up and spin-down transmission probabilities exhibit a qualitative symmetry under simultaneous spin reversal and energy inversion, closely following the relation $\mathcal{T}_\alpha^\uparrow(E) = \mathcal{T}_\alpha^\downarrow(-E)$.

Figure~\ref{Fig3} depicts the bias dependence of the spin-resolved currents along the three directions for the SRH helix, where panels (a), (b), and (c) correspond to the $x$-, $y$-, and $z$-axes, respectively. The calculations utilize the same parameter set specified in Fig.~\ref{Fig2} at a fixed Fermi energy $E_F=0.5\,$eV. Across all three directions, a distinct separation between the spin-up and spin-down currents is observed over a wide bias range. This different originates directly from the spin-channel asymmetry in the transmission spectra presented in Fig.~\ref{Fig2}. These unequal contributions signify the generation of a net spin-polarized current along all three spatial axes, confirming that the interplay between the non-collinear magnetic texture and SOC yields a truly multidirectional spin polarization. A comparison of Figs.~\ref{Fig3}(a)-(c) reveals that the splitting is most pronounced along the $z$-direction, indicating that the corresponding spin polarization component will be the largest. Conversely, the splittings along the $x$- and $y$-directions are notably smaller, reflecting the anisotropic nature of spin transport in the SRH helix. Furthermore, the current profiles exhibit distinct directional characteristics. For the $x$-direction, the spin-down current remains larger than its spin-up counterpart across most of the bias range, resulting in a predominantly negative spin polarization. In contrast, along the $y$- and $z$-directions, the spin-up and spin-down currents cross each other as the bias varies, indicating that these polarization components change sign as a function of the applied voltage.

Figure~\ref{Fig4} illustrates the bias dependence of the spin polarization coefficients along the three directions for the SRH helix, calculated using the same parameters as in Fig.~\ref{Fig2}. Several prominent features emerge from this analysis. First, a finite spin polarization is observed along all three spatial directions, confirming that the coexistence of a non-collinear magnetic texture and spin-orbit coupling generates a truly multidirectional spin polarization. This behavior aligns well with the spin-resolved transmission spectra and current-voltage characteristics discussed earlier. Second, the magnitude of the polarization depends strongly on the chosen spin quantization axis. Among the three components, $P_z$ exhibits the largest magnitude across most of the bias range, reaching nearly $35\%$ at low bias. In contrast, $P_x$ and $P_y$ remain comparatively smaller at this Fermi energy. Furthermore, the directional components show distinct behaviors regarding their signs. The $x$-component of the polarization remains predominantly negative across the entire bias range, whereas the $y$- and $z$-components undergo sign changes as the bias voltage is varied, in agreement with the current-voltage characteristics shown in Fig.~\ref{Fig3}.
\begin{figure*}[t!]
~~~~~~\includegraphics[width=0.3\textwidth]{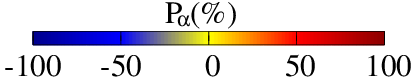}
\vskip 0.1 in
\includegraphics[width=0.32\textwidth]{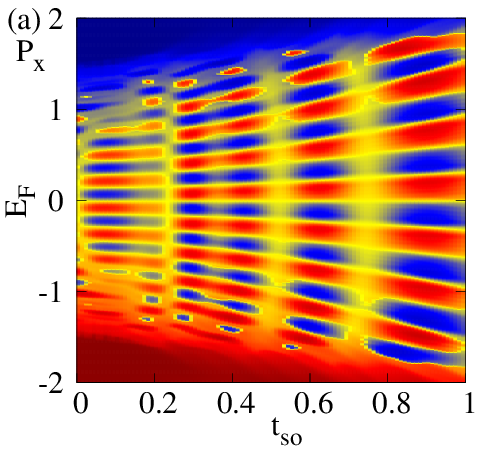}\hfill
\includegraphics[width=0.32\textwidth]{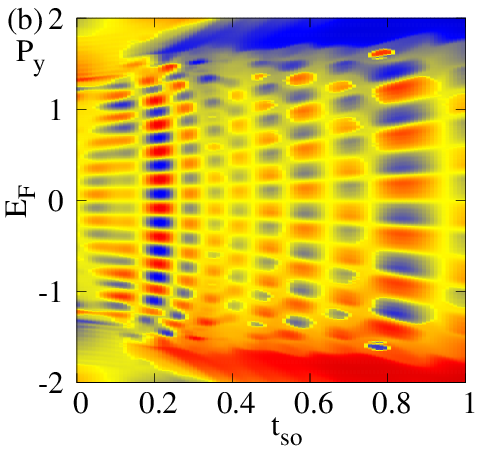} \hfill
\includegraphics[width=0.32\textwidth]{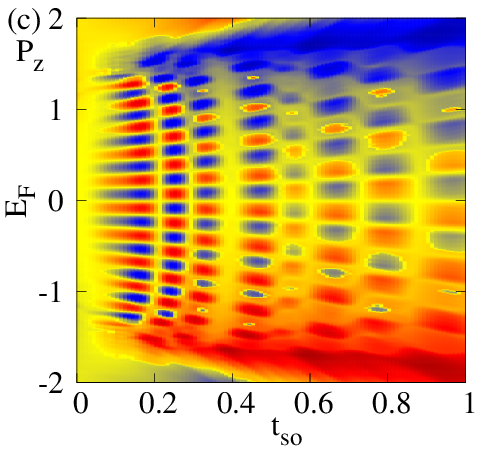}
\caption{(Color online). Maximum values of spin polarization obtained within the bias range $0$ to $1\,$V as functions of the SOC strength $t_{\rm so}$ and Fermi energy $E_F$ for the SRH helix. Panels (a), (b), and (c) show $P_x$, $P_y$, and $P_z$, respectively. Positive (red) and negative (blue) values correspond to opposite spin-polarization directions, as indicated by the color bar. The system parameters are the same as Fig.~\ref{Fig2}.}
\label{Fig5}
\end{figure*}


To obtain a comprehensive understanding of the spin-filtering characteristics, for the SRH configuration, we investigate the dependence of the spin polarization on both the Fermi energy and SOC strength. Figures~\ref{Fig5}(a-c) present the density plot for the maximum values of $P_x$, $P_y$, and $P_z$, respectively, obtained within the bias window $0-1\,$V, as functions of $E_F$ and $t_{\rm so}$. The dark-blue region means highly negative polarization whereas the dark-red regions means highly positive polarization. Several interesting features emerge from these density plots. 

First, all the three spin-polarization components exhibit strong variations with both $E_F$ and $t_{\rm so}$. For the $x$-component, even in the absence of SOC ($t_{\rm so} = 0$), a highly pronounced spin polarization survives at the outer energy bands, displaying strongly negative values (blue) at positive Fermi energies and strongly positive values (red) at negative Fermi energies. This provides direct evidence that the non-collinear helical antiferromagnetic texture alone breaks global parity-time-reversal (${\mathcal{PT}}$) symmetry, which is the fundamental prerequisite to induce intrinsic spin-polarized transport. A similar, though noticeably weaker, intrinsic signature is observed for the $y$-component at $t_{\rm so} = 0$ (Fig.~\ref{Fig5}(b)). As the SOC strength increases, the degree of polarization for $P_y$ also increases, particularly at higher Fermi energies on both the positive and negative sides of the band. In stark contrast, the $z$-component of the polarization ($P_z$) is identically zero at $t_{\rm so} = 0$. It becomes finite and achieves significant magnitudes, comparable to the $y$-component case, once a finite SOC is introduced.

The physical mechanism underlying the vanishing of $P_z$ at $t_{\rm so} = 0$ originates directly from the coplanar arrangement of the local magnetic moments and the resulting inherent antiunitary spin symmetry. In our helical configuration, the magnetic moment at the first site is aligned along the $x$-direction (see Fig.~\ref{fig1:diagram}), and the subsequent moments progressively rotate within the $xy$ plane by a total angle of $2\pi$. This rotation ensures a net zero magnetization across the helix ($\sum_{n=1}^N \mathbf{S}_n = 0$). Because the rotation is strictly confined to the plane, the out of plane component of every local moment is identically zero ($S_{n,z} = 0$). In the absence of SOC, the spin-independent hopping elements are purely real, while the complex off-diagonal onsite spin-flip terms exhibit a perfectly conjugate symmetric matrix structure. This structural arrangement ensures that the total Hamiltonian commutes with an antiunitary symmetry operator defined as $\Theta = \sigma_x K$, where $\sigma_x$ represents a spin-rotation operator and $K$ denotes the complex conjugation operation. The complex conjugation operation flips the phases of the complex spin-flip terms, and the subsequent spin rotation physically restores them to their original matrix positions, leaving the entire Hamiltonian invariant. Since $\Theta \sigma_z \Theta^{-1} = -\sigma_z$, it forces the spin up and spin down transmission channels along the $z$-direction to be perfectly balanced. Consequently, we have $P_z = 0$. Such a vanishing spin polarization of a particular component is consistent with the spin-group symmetry rules that forbid specific spin polarization components when spin and spatial degrees of freedom are decoupled~\cite{liu2022}. Once SOC is included, the antiunitary symmetry is broken. This further, lifts the constraints on the longitudinal channel, mixing the out-of-plane spin states with the in-plane transport channels and $P_z$ becomes finite. 

The finite $P_x$ and $P_y$ in the absence of SOC can also be understood in terms of this antiunitary symmetry operation. We find that the symmetry transformations yield $\Theta \sigma_x \Theta^{-1} = +\sigma_x$ and $\Theta \sigma_y \Theta^{-1} = +\sigma_y$. Because $\Theta$ commutes with both transverse spin operators, this antiunitary symmetry imposes no parity constraints on the transverse transport channels. Therefore, the broken global parity-time reversal symmetry generated exclusively by the non-collinear helical magnetic texture is capable of achieving non-zero $P_x$ and $P_y$ even in the absence of SOC~\cite{gurung}.

Furthermore, the specific spatial parity of the in-plane magnetic texture explains why the intrinsic polarization along the $x$-axis is significantly larger than that along the $y$-axis. Although the net sums of both the $x$- and $y$-components of the local moments vanish independently over the entire helix, their local configurations possess distinct geometric symmetries. Due to the initial alignment along the $x$-axis, the $y$-components, which vary as $\sin\varphi_n$, possess an odd spatial parity under reflection across the center of the helix. Consequently, every positive local $y$-component finds a matching negative counterpart at a symmetric site index, resulting in a mutual destructive interference of the local spin-dependent scattering fields that protects the channel and heavily suppresses the net spin transport along the $y$-axis. In contrast, the $x$-components, which vary as $\cos\varphi_n$, exhibit an even spatial parity and do not possess this site-by-site opposite pairing. This structural difference allows the spin-dependent scattering from the $x$-components to accumulate constructively, leaving the $x$-spin transport channel fully exposed to the broken inversion symmetry of the chiral helix, which results in a much stronger intrinsic spin polarization along the $x$-axis compared to the transverse $y$-axis.

Another interesting feature we observe in the density plots is the fringe structures characterized by alternating zones of positive (red) and negative (blue) spin polarization. This phenomenon can be understood from the spin-resolved transmission spectra (Fig.~\ref{Fig2}). The outer energy windows exhibit strong unidirectional polarization, where either the spin-up (red) or spin-down (black) channel achieves near-perfect transmission while its counterpart is blocked. On the other hand, the central energy band spanning from $E \approx -1.5$ to $1.5$ displays a dense series of nearly equidistant resonance peaks for both spin channels. These transmission peaks are systematically out of phase and interleaved with one another. As the energy varies continuously through this central regime, the dominant transmission channel rapidly alternates between spin-up and spin-down channels. Because the net polarization satisfies $P_\alpha \propto \mathcal{T}_\uparrow - \mathcal{T}_\downarrow$, this comb-like staggering of the resonance peaks forces the spin polarization to fluctuate periodically between positive and negative values, thereby generating the alternating red and blue fringe patterns.

The above results establish that the SRH helix supports finite spin polarization along all three spatial directions. The corresponding polarization components, however, exhibit distinct magnitudes and bias dependences, reflecting anisotropic spin transport. It is therefore natural to ask how these characteristics are modified when long-range hopping is incorporated. Since the LRH geometry allows electrons to propagate through additional hopping paths, and its interplay with the non-collinear magnetic texture and SOC can differ substantially from the SRH case. We now turn to the LRH configuration to investigate how the inclusion of long-range hopping influences these spin-transport properties.

\begin{figure*}[t!]
\includegraphics[width=0.32\textwidth]{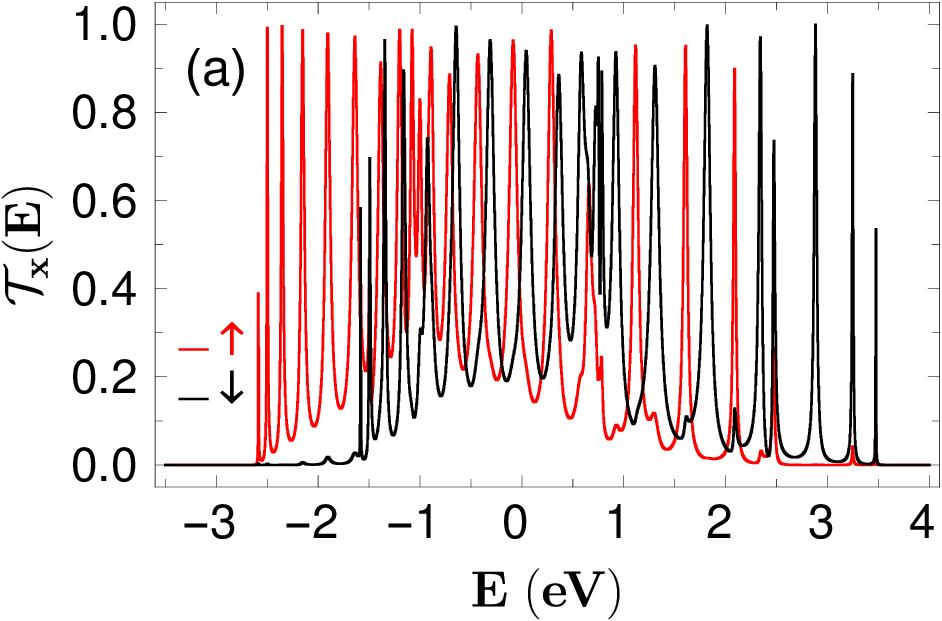}\hfill
\includegraphics[width=0.32\textwidth]{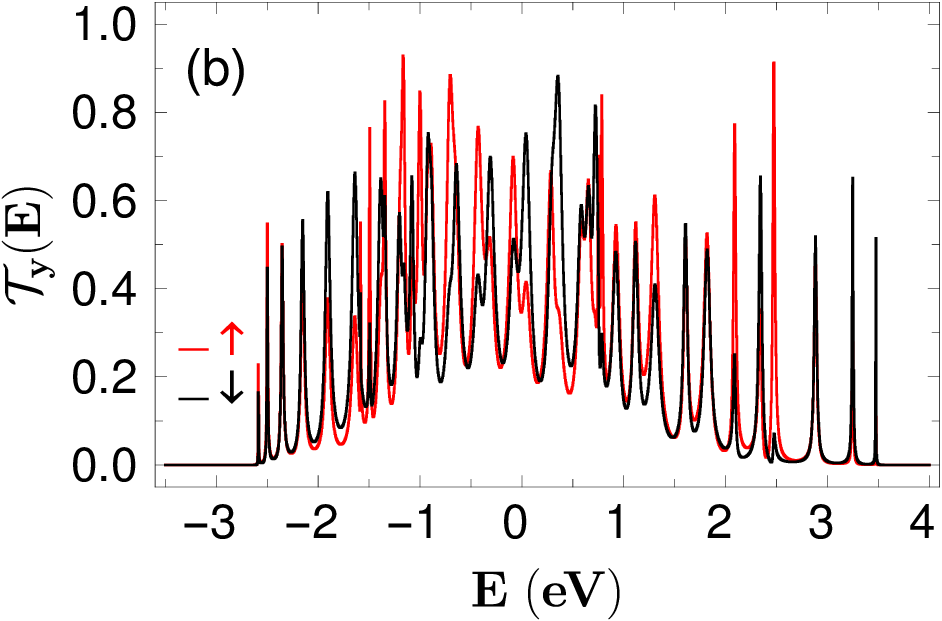}\hfill
\includegraphics[width=0.32\textwidth]{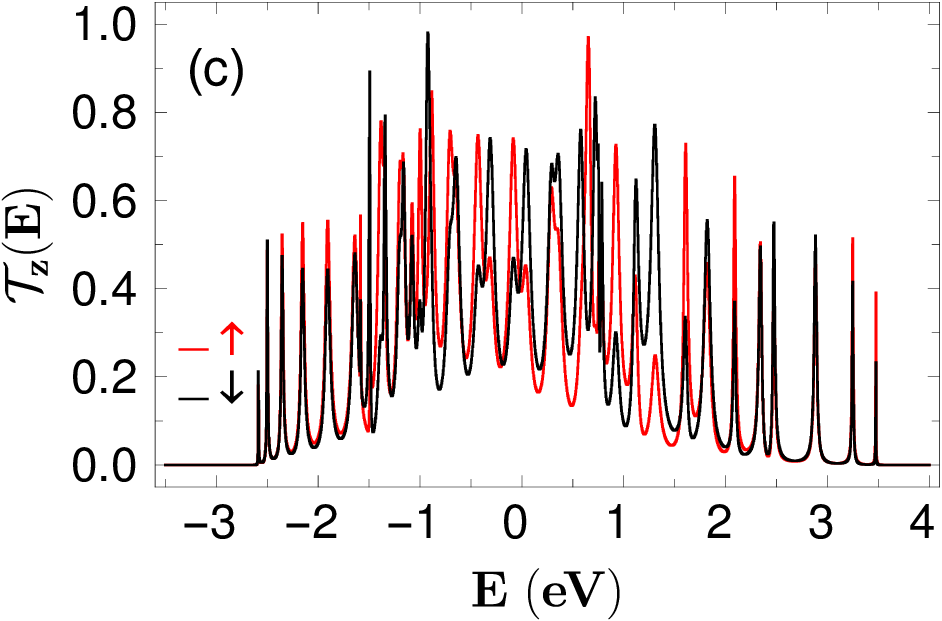}
\caption{(Color online). Spin-resolved transmission probabilities ${\mathcal{T}}_\alpha^{\uparrow(\downarrow)}$ $(\alpha=x,y,z)$ as a function of energy for the LRH helix. (a) $x$-, (b) $y$-, and (c) $z$-directions. The red and black curves correspond to the up- and down-spin transmission channels. The calculations are carried out for $N=20$, $t_{\rm so}=0.25$, and $\mathpzc{h}=0.5$.}
\label{Fig6}
\end{figure*}
\begin{figure*}[ht!]
\includegraphics[width=0.3\textwidth]{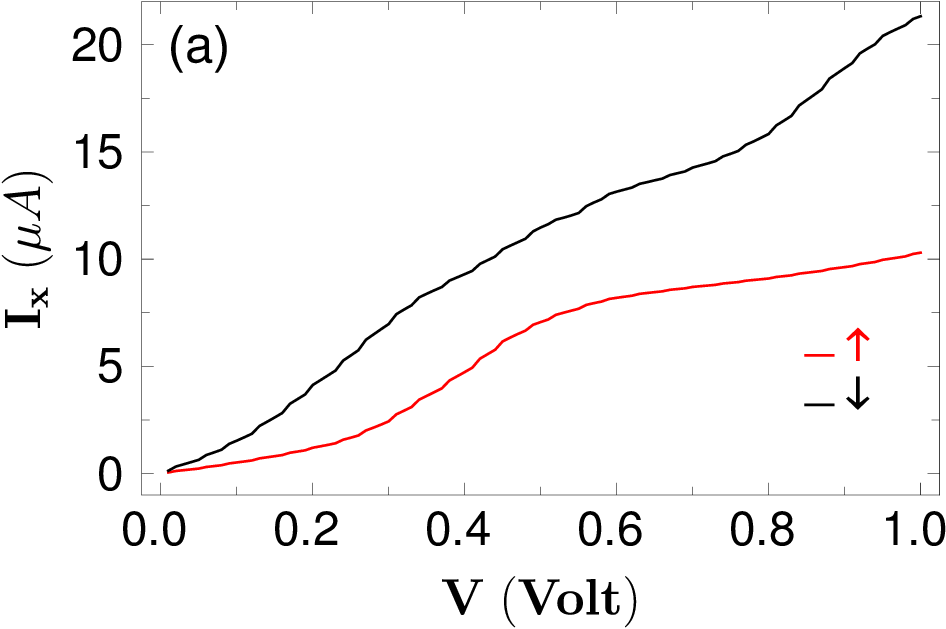}\hfill
\includegraphics[width=0.3\textwidth]{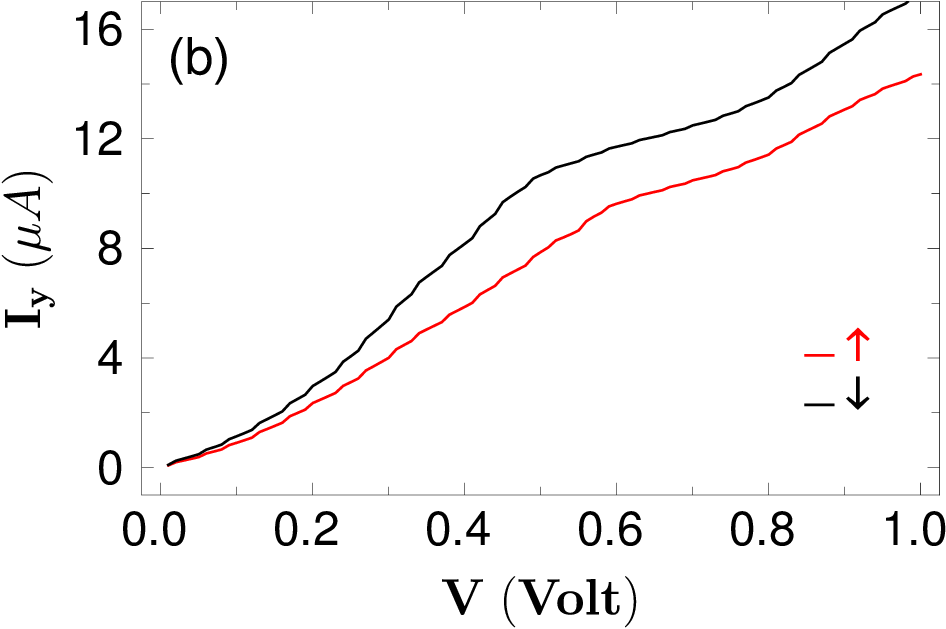}\hfill
\includegraphics[width=0.3\textwidth]{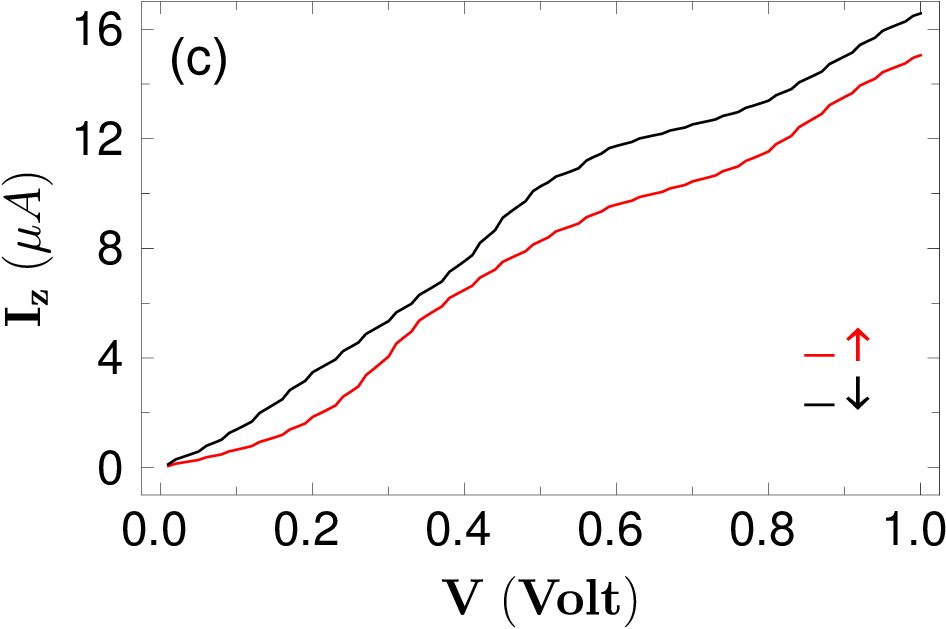}
\caption{(Color online). Bias dependence of the spin-resolved current $I_\alpha^{\uparrow(\downarrow)}$ $(\alpha=x,y,z)$ for the LRH configuration. Panels (a), (b), and (c) show the currents along the $x$-, $y$-, and $z$-directions, respectively. The red (black) curves correspond to the up- and down-spin currents, respectively. The spin currents are computed at a fixed Fermi energy $E_F=0.5\,$eV. All the system parameters are the same as Fig.~\ref{Fig6}.}
\label{Fig7}
\end{figure*}
\begin{figure*}[t!]
\includegraphics[width=0.32\textwidth]{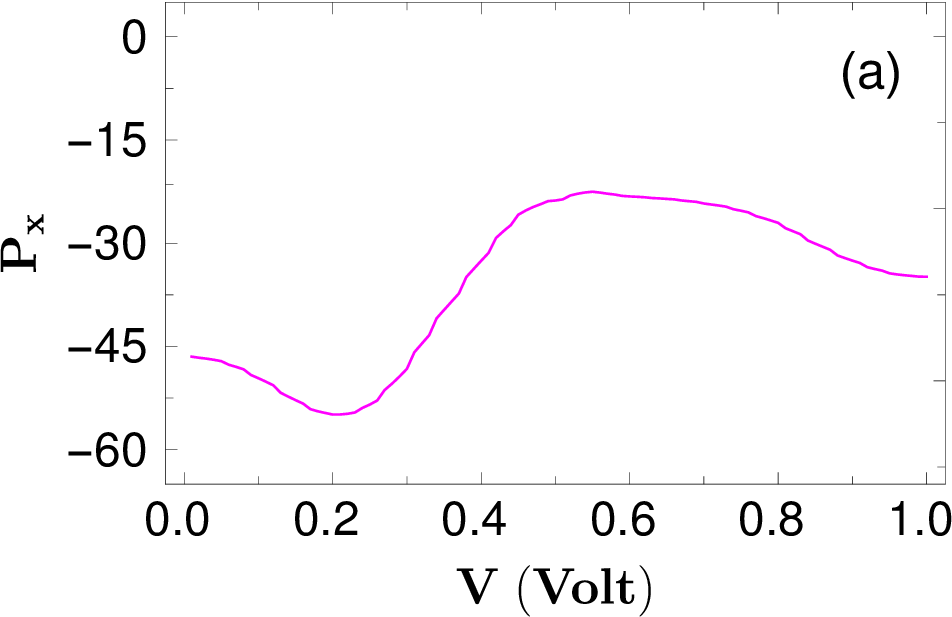}\hfill
\includegraphics[width=0.32\textwidth]{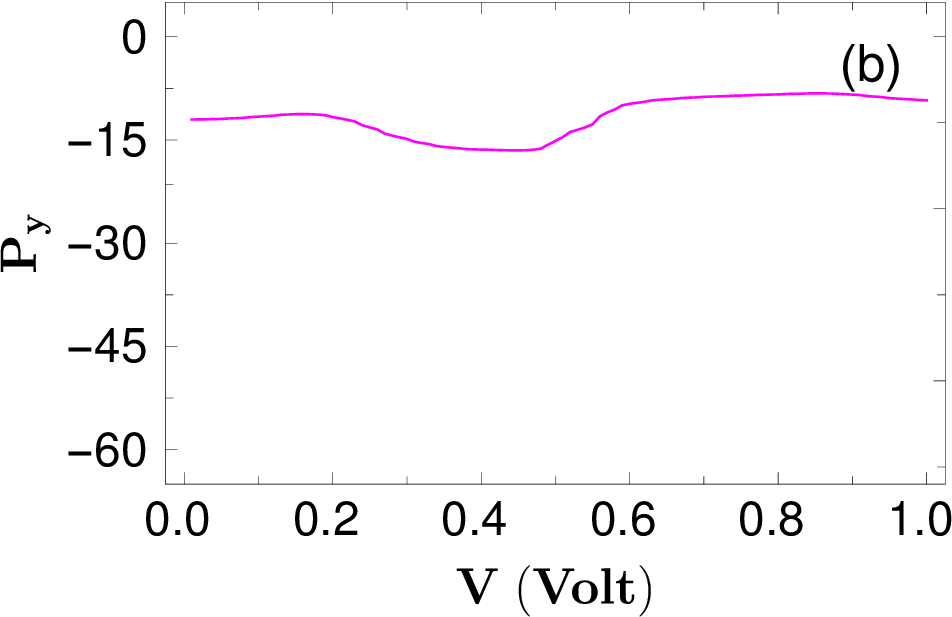}\hfill
\includegraphics[width=0.32\textwidth]{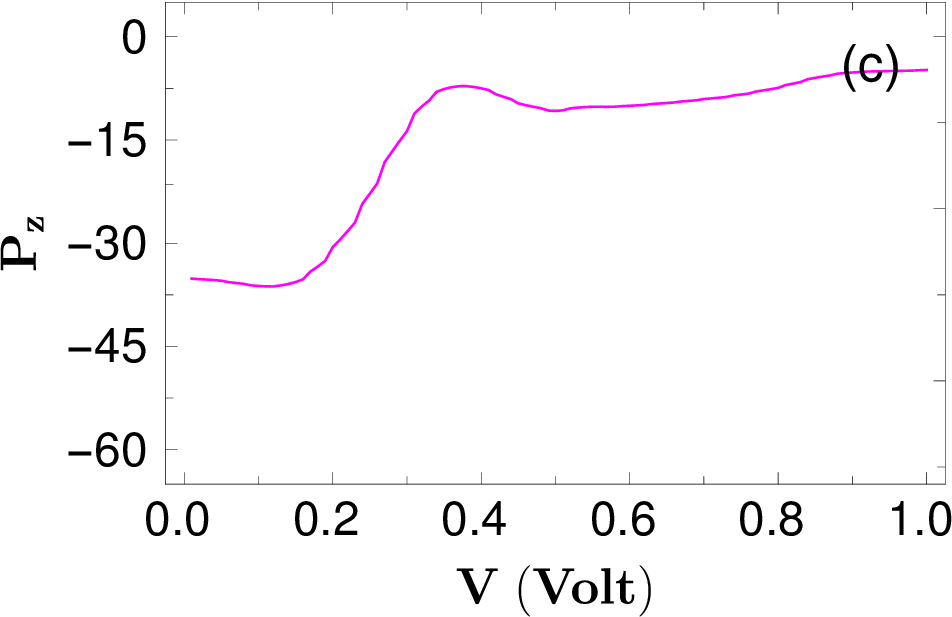}
\caption{(Color online). Bias dependence of the spin polarization coefficient $P_\alpha$ $(\alpha=x,y,z)$ for the LRH configuration. Panels (a), (b), and (c) show the spin polarization coefficients $P_x$, $P_y$, and $P_z$, respectively, as a function of the applied bias voltage. Fermi energy is set at $E_F=0.5\,$eV and all the system parameters are the same as Fig.~\ref{Fig6}.}
\label{Fig8}
\end{figure*}


\subsubsection{Long-range hopping scenario }

Figure~\ref{Fig6} displays the behavior of the spin-resolved transmission probabilities as a function of energy for the LRH case. Panels (a)-(c) correspond to the $x$-, $y$-, and $z$-directions, respectively, with the red (black) curves representing the spin-up (spin-down) transmission probabilities. The calculations are performed for $N=20$, $t_{\rm so}=0.25$, and $\mathpzc{h}=0.5$. Compared to the SRH case, the LRH configuration yields a distinctly altered transmission profile. The spin-up and spin-down transmission probabilities remain strongly mismatched over a broad energy range for all three spin directions. Crucially, this spin-channel asymmetry is no longer restricted to a narrow energy band. Instead of that, substantial deviations between the two channels persist across the entire allowed energy window. Furthermore, the transmission resonances exhibit a highly non-uniform spectral distribution, where the peaks are densely packed within the lower energy regime but become progressively sparse and widely separated as the energy increases. This variable peak density is a characteristic signature of transport networks governed by long-range coupling interactions~\cite{dey2023,ganguly2026}. Mathematically, this spectral irregularity directly reflects a fundamental breakdown of electron-hole symmetry. Because the extended spatial range of electron hopping heavily distorts the underlying energy dispersion, the energy intervals separating successive quantum resonant states widen progressively in the high-energy domain, naturally giving rise to the visible expansion of the peak separation.

\begin{figure*}[t!]
~~~~~~\includegraphics[width=0.3\textwidth]{cb.eps}\vskip 0.1 in
\includegraphics[width=0.32\textwidth]{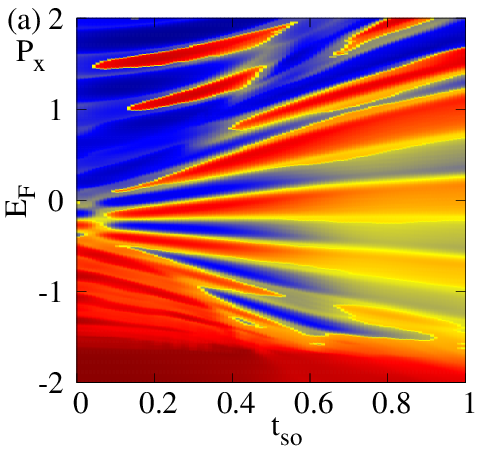}\hfill
\includegraphics[width=0.32\textwidth]{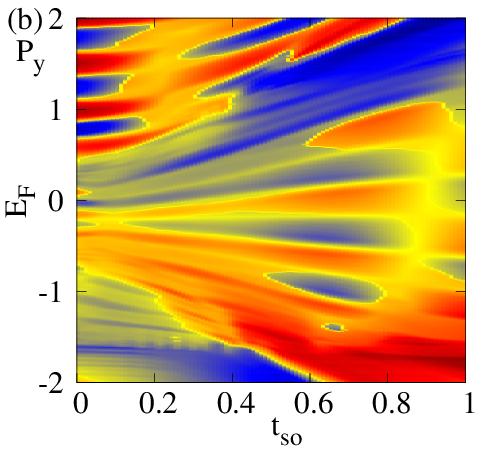}\hfill
\includegraphics[width=0.32\textwidth]{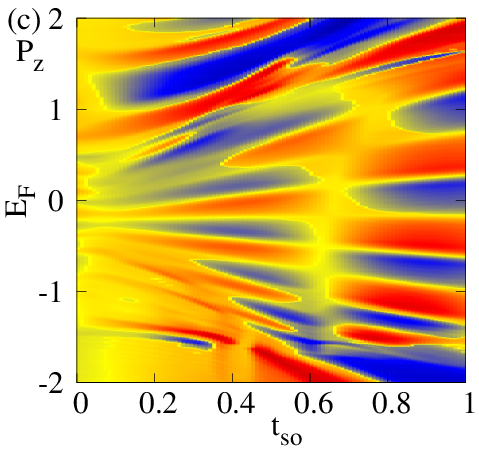}
\caption{(Color online). Maximum values of spin polarization coefficient obtained within the bias range $0$ to $1\,$V as functions of the SOC strength $t_{\rm so}$ and Fermi energy $E_F$ for the LRH helix. Panels (a), (b), and (c) show $P_x$, $P_y$, and $P_z$, respectively. Colorbar has the same convention as Fig.~\ref{Fig5}. All the system parameters are the same as Fig.~\ref{Fig6}.}
\label{Fig9}
\end{figure*}
Along the $x$-direction, a strong separation between the two spin channels is visible over a wide energy interval, as shown in Fig.~\ref{Fig6}(a). A similar behavior is found along the $y$-direction (Fig.~\ref{Fig6}(b)), where the transmission peaks of the two spin channels differ substantially in magnitude and appear at different energy positions. The $z$-component likewise exhibits a pronounced spin-channel imbalance across the transmission band, as shown in Fig.~\ref{Fig6}(c). These results indicate that the LRH geometry supports a significant spin polarization along all the three directions.

It is apparent that the LRH case yields a far more effective spin-splitting effect than the SRH configuration. The inclusion of these extended hopping channels radically modifies the quantum interference landscapes of the propagating electrons, which dramatically strengthens their concurrent interplay with the non-collinear magnetic texture and the SOC. Consequently, the resulting spin-splitting becomes remarkably robust across the spectrum, creating exceptionally favorable conditions for the generation of multidirectional spin polarization. This assertion will be verified explicitly by the transport current and spin-polarization calculations presented below.

Figure~\ref{Fig7} presents the bias dependence of the spin-resolved currents for the LRH helix, calculated using the parameters mentioned in Fig.~\ref{Fig6}. Panels (a)-(c) correspond to the $x$-, $y$-, and $z$-directions, respectively, with the currents evaluated at a fixed Fermi energy $E_F=0.5\,$eV. In contrast to the SRH case, a pronounced separation between the spin-up and spin-down currents is observed along all three directions. The most significant current splitting occurs along the $x$-direction, where the spin-down current remains considerably larger than the spin-up current over the entire bias range. Consequently, a large negative spin polarization is expected along this axis. Along the $y$-direction, the two spin currents are comparable at low bias but their difference increases as the bias increases, yielding a finite spin polarization. A substantial current splitting is likewise observed along the $z$-direction, where the spin-down current again dominates across most of the bias range.

Figure~\ref{Fig8} depicts the behavior of the spin polarization coefficient $P_\alpha$ for the LRH case, where panels (a), (b), and (c) display $P_x$, $P_y$, and $P_z$, respectively. The results are obtained using the same parameter set as the preceding current calculations. Unlike the SRH configuration, the LRH helix exhibits an appreciable spin polarization along all three directions over the entire bias range. The polarization remains robustly finite for all applied voltages, demonstrating efficient multidirectional spin filtering. The $x$-component, shown in Fig.~\ref{Fig8}(a), remains negative throughout the bias range. Its magnitude increases rapidly at low bias, reaches a maximum of approximately $50\%$ around $0.2\,$V, and then decreases slightly at higher voltages while remaining strongly negative. The $y$-component (Fig.~\ref{Fig8}(b)) exhibits a comparatively smaller magnitude of roughly $20\%$, remaining negative across the entire bias window with only moderate variation. In contrast, the $z$-component (Fig.~\ref{Fig8}(c)) shows a significant spin polarization of up to $40\%$ at low bias, which steadily decreases with increasing voltage and approaches zero at higher bias limits.

These results differ significantly from those obtained for the SRH helix, directly validating the expectations drawn from the transmission spectra. The above findings confirm that the LRH helix provides a significantly more effective platform for generating multidirectional spin-polarized currents.

\begin{figure*}[ht!]
\centering
\includegraphics[width=0.32\textwidth]{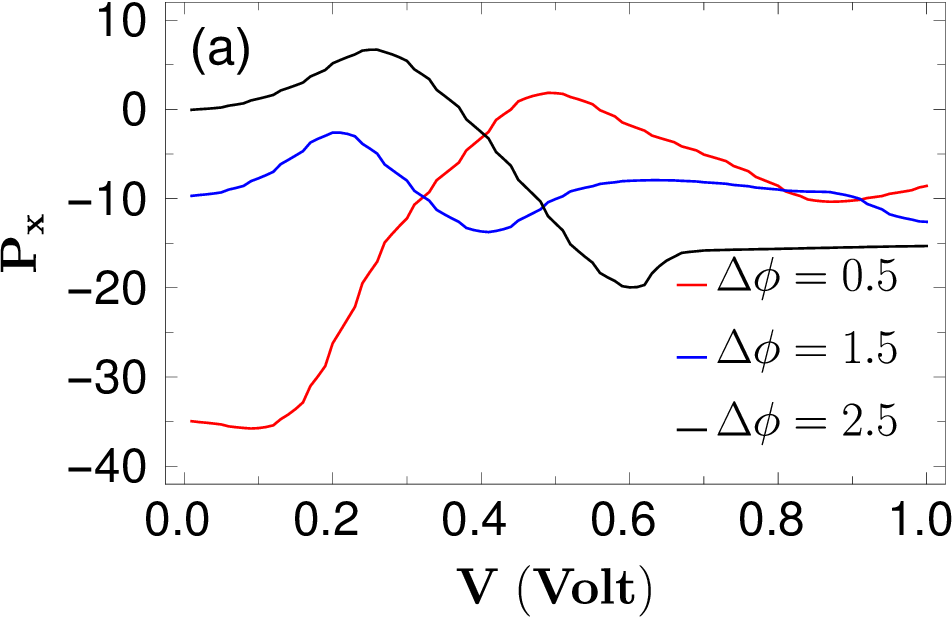} \hspace{0.1cm}
\includegraphics[width=0.32\textwidth]{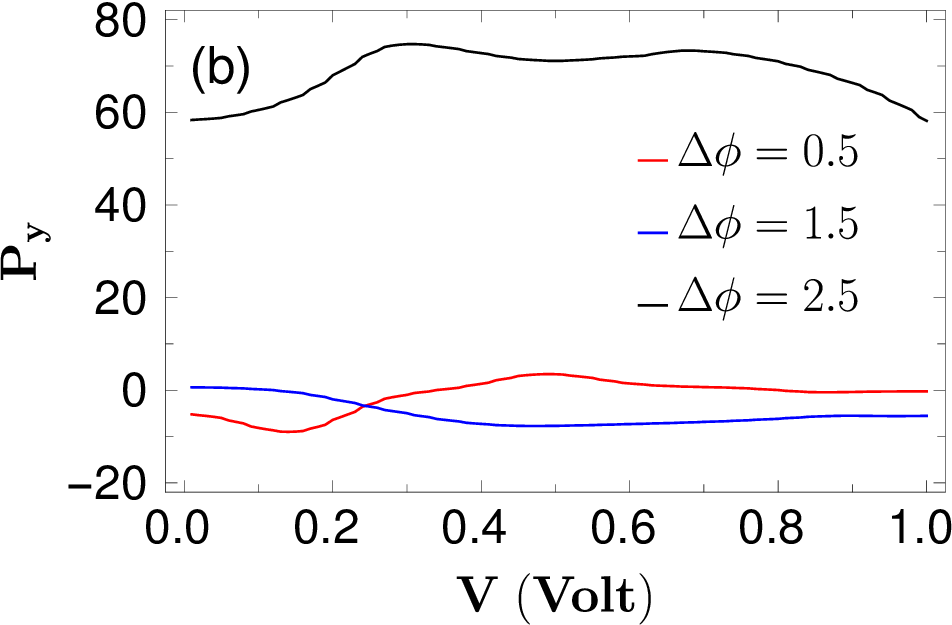} \hspace{0.1cm}
\includegraphics[width=0.32\textwidth]{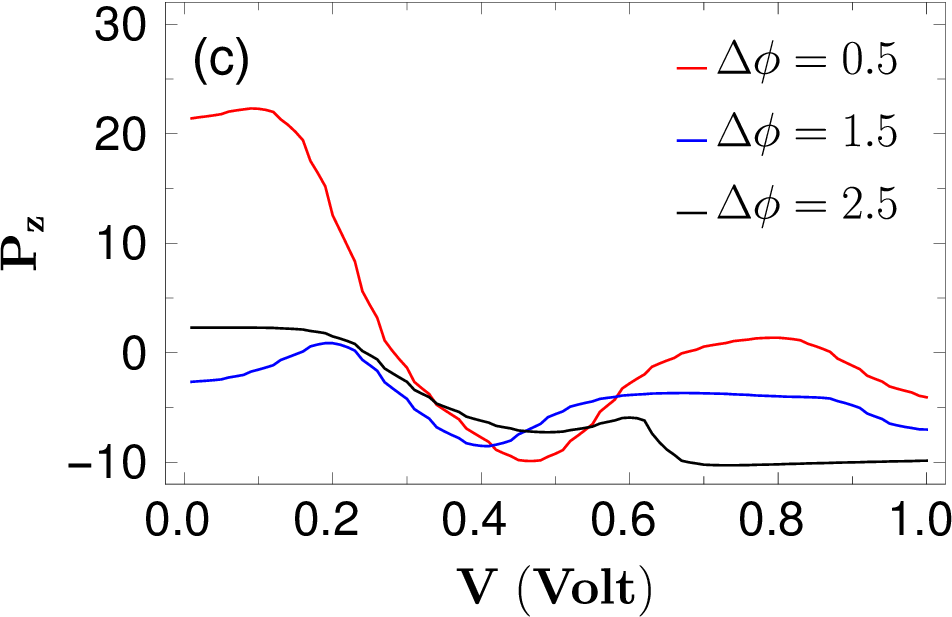} \hspace{0.1cm}
\caption{(Color online). Bias dependence of the spin polarization components for different helical twist angles $\Delta\phi$ in the LRH helix with $N=20$, $t_{\rm so}=0.25$, $h=0.5$, and $E_F=0.5\,$eV. Panels (a), (b), and (c) display $P_x$, $P_y$, and $P_z$, respectively. The red, blue, and black curves correspond to $\Delta\phi=0.5$, $1.5$, and $2.5$, respectively.}
\label{Fig10}
\end{figure*}

To gain a broader understanding of the role of SOC and Fermi energy on the spin-filtering performance, we map the maximum spin polarization obtained within the bias range $0$ to $1\,$V across the $(t_{\rm so},E_F)$ parameter space. The results for the LRH helix are presented in Fig.~\ref{Fig9}, where panels (a)-(c) correspond to $P_x$, $P_y$, and $P_z$, respectively. Several important features emerge from these density plots. First, a robust spin polarization is observed across a wide range of both $t_{\rm so}$ and $E_F$, indicating that the multidirectional spin-filtering effect is not restricted to a finely tuned parameter regime. Instead, large positive and negative polarization values persist throughout extended domains of the parameter space. Second, the magnitude and sign of the spin polarization exhibit a high sensitivity to the concurrent variations of $t_{\rm so}$ and $E_F$. Clear, alternating bands of positive and negative polarization span all three panels, separated by crossover regions where the polarization changes sign. These prominent sign reversals demonstrate that the dominant transport spin channel can be actively switched by adjusting either the SOC strength or the Fermi energy. This provides a convenient route for the electrical control of the spin-polarization direction. A comparative analysis among the three polarization components reveals notable differences. The $x$-component, illustrated in Fig.~\ref{Fig9}(a), displays the most pronounced overall magnitude, with the polarization efficiency approaching nearly $100\%$ within extended parameter blocks. The $y$- and $z$-components likewise display substantial polarization, reaching values of approximately $75\%$ while exhibiting similarly rich landscapes of sign reversal, as observed in Figs.\ref{Fig9}(b) and (c), respectively. Remarkably, at $t_{\rm so} = 0$, $P_z$ vanishes identically across all Fermi energies due to the antiunitary spin symmetry mentioned earlier in the SRH case, whereas the transverse components $P_x$ and $P_y$ remain finite. The occurrence of large polarization amplitudes along all three Cartesian directions further confirms that the LRH configuration supports a genuine multidirectional spin filtering effect. Ultimately, Fig.~\ref{Fig9} demonstrates that the synergetic action of long-range hopping, non-collinear magnetic ordering, and SOC produces a highly robust and tunable multidirectional spin polarization over a wide parameter range.


\begin{figure*}[t!]
\centering
\includegraphics[width=0.32\textwidth]{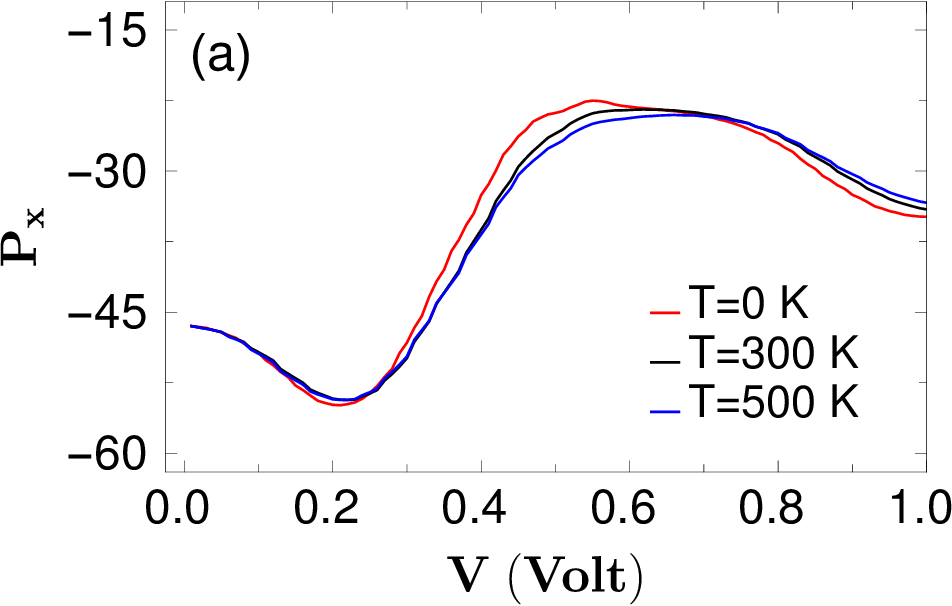} \hspace{0.1cm}
\includegraphics[width=0.32\textwidth]{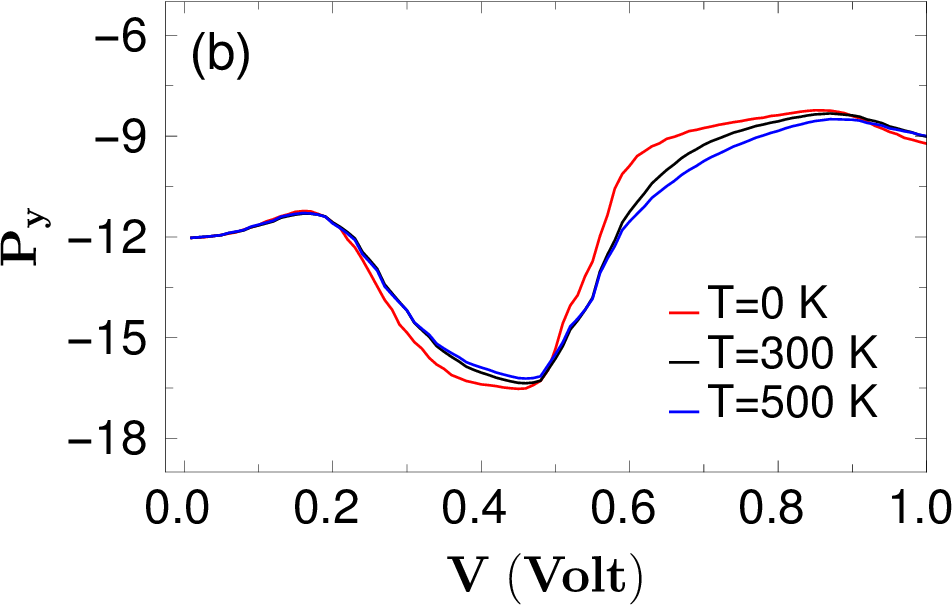} \hspace{0.1cm}
\includegraphics[width=0.32\textwidth]{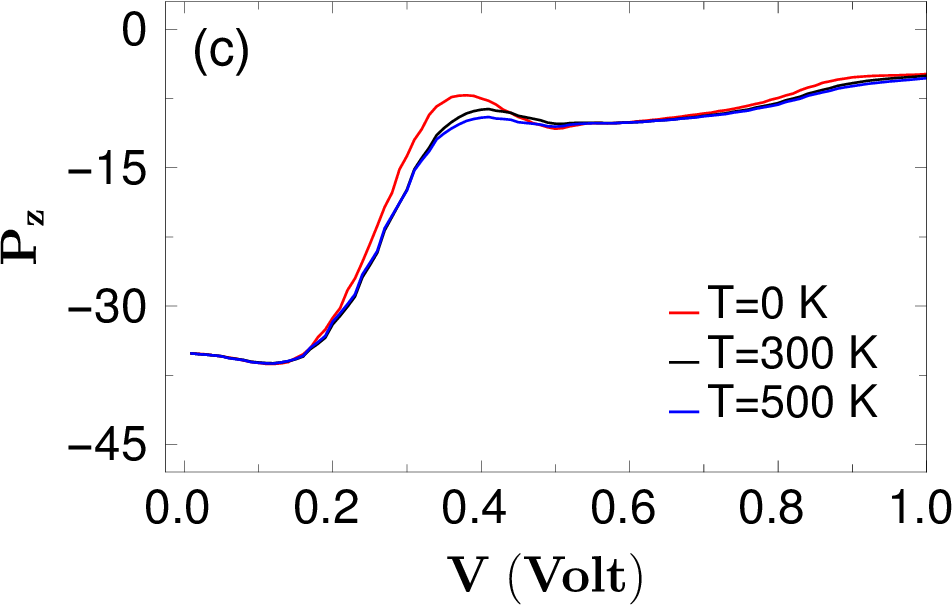} \hspace{0.1cm}
\caption{(Color online). Temperature dependence of the spin polarization coefficients for the LRH helix with $N=20$, $t_{\rm so}=0.25$, $h=0.5$, and $E_F=0.5\,$eV. Panels (a), (b), and (c) display the bias dependence of $P_x$, $P_y$, and $P_z$, respectively, for three representative temperatures: $T=0$, $300$, and $500\,$K.}
\label{Fig11}
\end{figure*}
\subsubsection{Effect of helicity on spin polarization }
To examine the influence of the helical geometry on spin-selective transport, we investigate the bias voltage dependence of the spin polarization for distinct values of the twisting angle $\Delta\phi$. The corresponding profiles are presented in Fig.~\ref{Fig10}, where panels (a)-(c) depict the polarization components $P_x$, $P_y$, and $P_z$, respectively. We consider three representative values for the twisting angle, namely, $\Delta\phi=0.5$, $1.5$, and $2.5\,$rad, while keeping the stacking distance and helix radius identical to the LRH configuration. The numerical results clearly demonstrate that the spin polarization is highly sensitive to the twisting angle of the helix. Notably, variations in $\Delta\phi$ induce dramatic modifications not only in the polarization magnitude but also in its sign.

The $x$-component of polarization, illustrated in Fig.~\ref{Fig10}(a), exhibits a profound dependence on $\Delta\phi$. At a small twisting angle ($\Delta\phi=0.5$), the polarization undergoes a clear sign reversal from negative to positive at intermediate bias voltages before declining at higher bias. For larger twisting angles, the polarization profiles remain predominantly negative, though their magnitudes vary significantly across the entire voltage window. The response of the $y$-component is even more promising. As demonstrated in Fig.~\ref{Fig10}(b), increasing the twisting angle enhances the spin polarization dramatically. While $P_y$ remains suppressed near zero for $\Delta\phi=0.5$ and $1.5$, it becomes remarkably large and positive for $\Delta\phi=2.5$ across the entire bias range, peaking near $75\%$. 

The $z$-component also undergoes substantial alterations with varying $\Delta\phi$, as shown in Fig.~\ref{Fig10}(c). For $\Delta\phi=0.5$, the polarization is notably positive at low bias and transitions into the negative regime as the bias voltage increases. In contrast, for larger twisting angles, $P_z$ remains relatively small and predominantly negative over the majority of the bias range. Overall, these findings demonstrate that the twisting angle provides a powerful mechanism for controlling both the sign and magnitude of the spin polarization components. Because tuning $\Delta\phi$ geometrically modifies the network of long-range hopping paths, it directly modulates the spin-dependent quantum interference pathways. Consequently, the structural chirality of the helix serves as an external tuning parameter for optimizing and customizing multidirectional spin polarization.

\subsubsection{Thermal stability of spin polarization }
To inspect the robustness of the spin-filtering effect against thermal fluctuations, we investigate the temperature dependence of the spin polarization components. Figure~\ref{Fig11} shows the bias dependence of $P_x$, $P_y$, and $P_z$ for the LRH helix at three different temperatures, namely $T=0$, $300$, and $500\,$K. The calculations are performed with $N=20$, $t_{\rm so}=0.25$, $\mathpzc{h}=0.5$, and $E_F=0.5\,$eV. Most important  feature of Fig.~\ref{Fig11} is the weak temperature dependence of all three polarization components. The magnitudes of $P_x$, $P_y$, and $P_z$ exhibit only minor quantitative variations, while their characteristic bias-dependent trends are preserved over the entire voltage range. The weak sensitivity of the polarization to temperature can be understood from the fact that the spin-selective transport originates from sizable differences between the spin-dependent transmission channels over relatively broad energy intervals. Consequently, thermal fluctuation does not substantially wash out the spin asymmetry, even at room temperature and beyond. These results demonstrate that the multidirectional spin polarization generated in the LRH helix is remarkably robust against thermal effects. Even at room temperature and above, the spin polarization retains nearly the same magnitude and bias dependence as that obtained at zero temperature. This thermal stability highlights the potential of the proposed system for practical spintronic applications operating under ambient conditions.


\subsubsection{System size dependence}
\begin{figure*}[ht]
\includegraphics[width=0.33\textwidth]{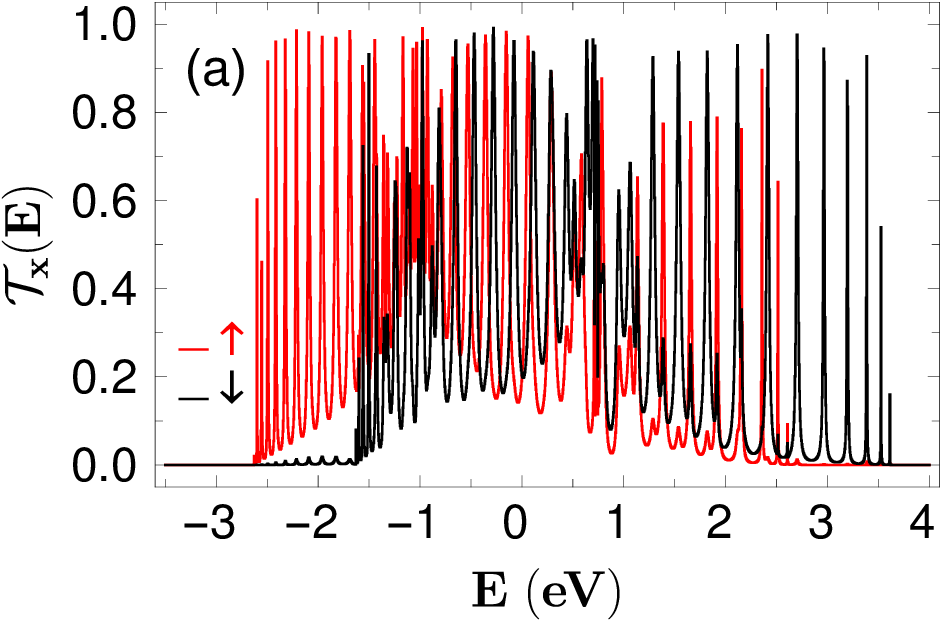}\hfill
\includegraphics[width=0.33\textwidth]{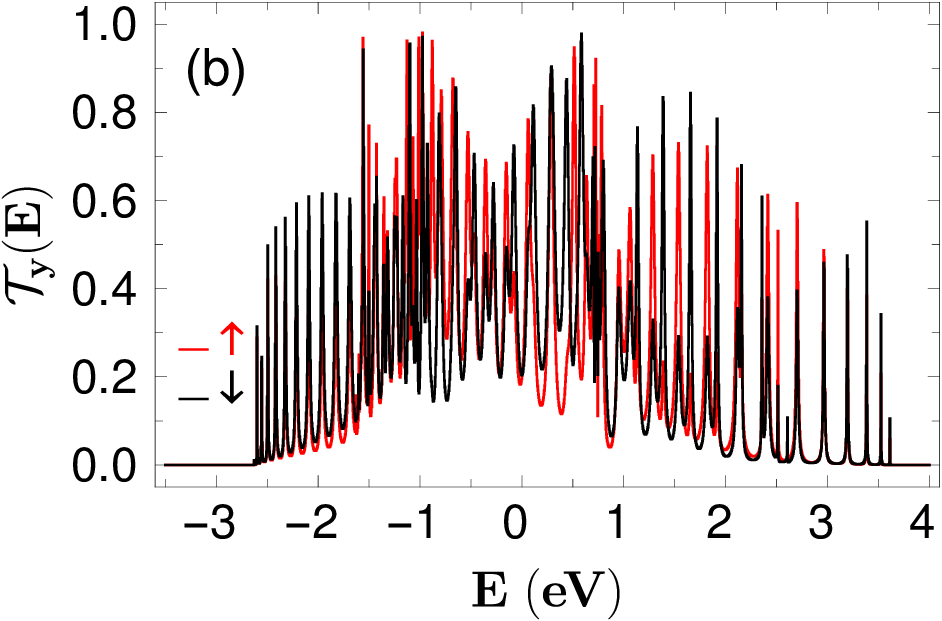}\hfill
\includegraphics[width=0.33\textwidth]{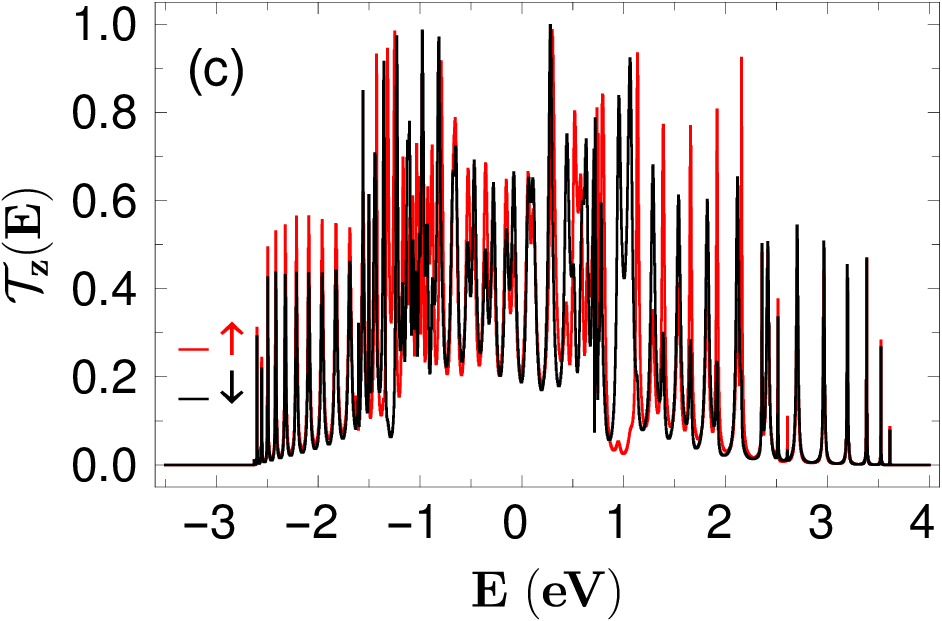}
\caption{(Color online). Spin-resolved transmission probabilities ${\mathcal{T}}_\alpha^{\uparrow(\downarrow)}$ $(\alpha=x,y,z)$ as a function of energy for the LRH helix. (a) $x$-, (b) $y$-, and (c) $z$-directions. The red and black curves correspond to the up- and down-spin transmission channels. The calculations are carried out for $N=40$, $t_{\rm so}=0.25$, and $\mathpzc{h}=0.5$.}
\label{fig12}
\end{figure*}

\begin{figure*}[ht!]
\includegraphics[width=0.3\textwidth]{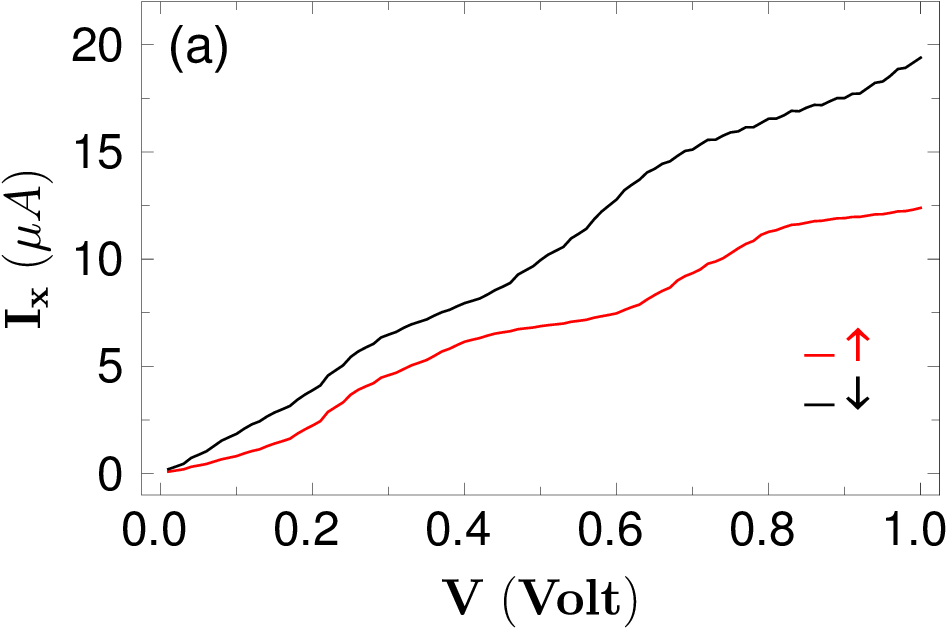}\hfill
\includegraphics[width=0.3\textwidth]{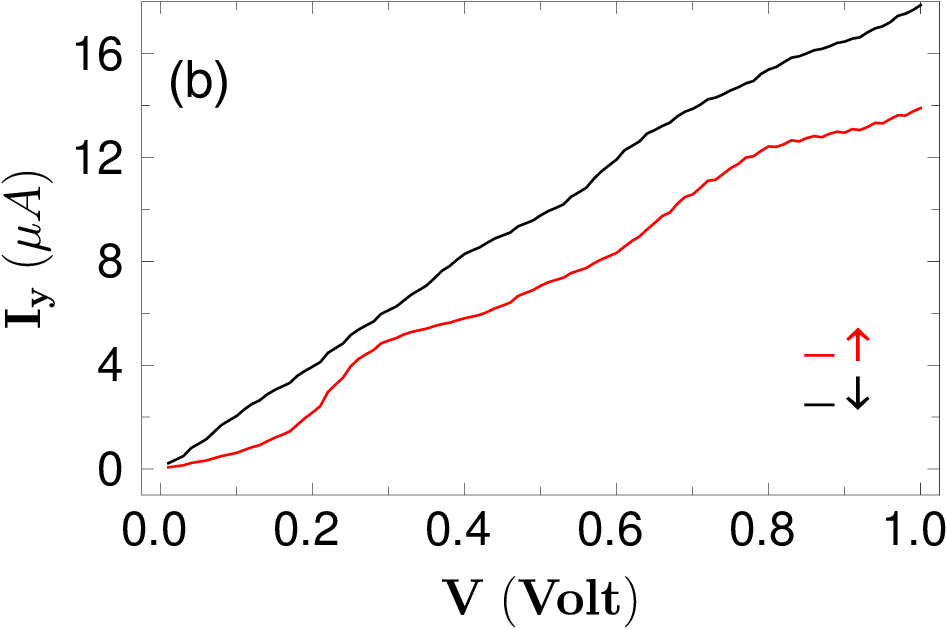}\hfill
\includegraphics[width=0.3\textwidth]{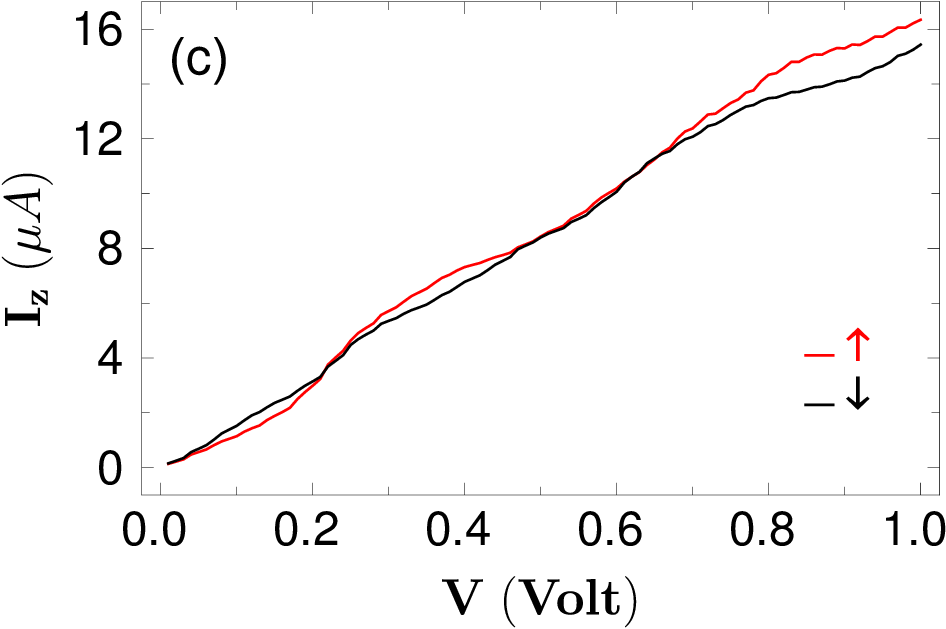}
\caption{(Color online). Bias dependence of the spin-resolved current $I_\alpha^{\uparrow(\downarrow)}$ $(\alpha=x,y,z)$ for the LRH configuration. Panels (a), (b), and (c) show the currents along the $x$-, $y$-, and $z$-directions, respectively. The red (black) curves correspond to the up- and down-spin currents, respectively. The spin currents are computed at a fixed Fermi energy $E_F=0.4\,$eV and for $N=40$.}
\label{fig13}
\end{figure*}

To investigate the effect of system size on the transport behavior, we extend our numerical calculations to larger LRH systems. Figure~\ref{fig12} displays the spin-resolved transmission probabilities ($\mathcal{T}_x, \mathcal{T}_y, \mathcal{T}_z$) as a function of energy for a system size $N = 40$. Because the number of allowed energy levels in the central region increases with the system size, a highly dense, oscillatory spectrum of resonant transmission peaks occurs spanning the allowed energy window. To extract the best spin-filtering performance for each structural configuration, the Fermi energy is independently adjusted to optimize the spin response. For $N=40$, we set $E_F=0.4\,$eV. The corresponding spin-resolved currents ($I_x, I_y, I_z$) for $N=40$, evaluated at this optimized Fermi energy ($E_F=0.4\,$eV), are presented in Fig.~\ref{fig13}. These currents increase steadily with the applied bias voltage and demonstrate that the fundamental multidirectional spin-filtering effect is robustly preserved.

\begin{figure*}[t!]
\includegraphics[width=0.32\textwidth]{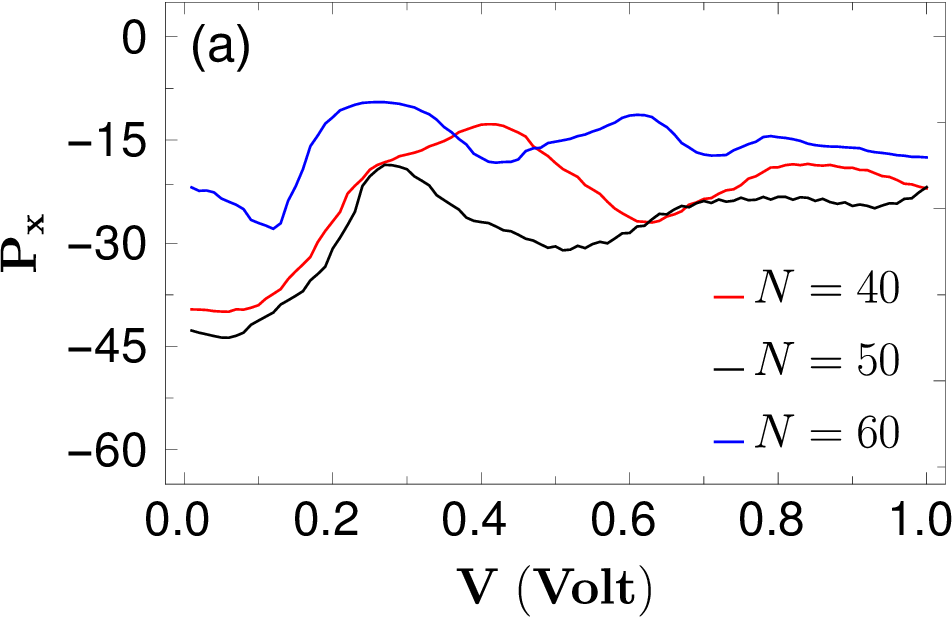}\hfill
\includegraphics[width=0.32\textwidth]{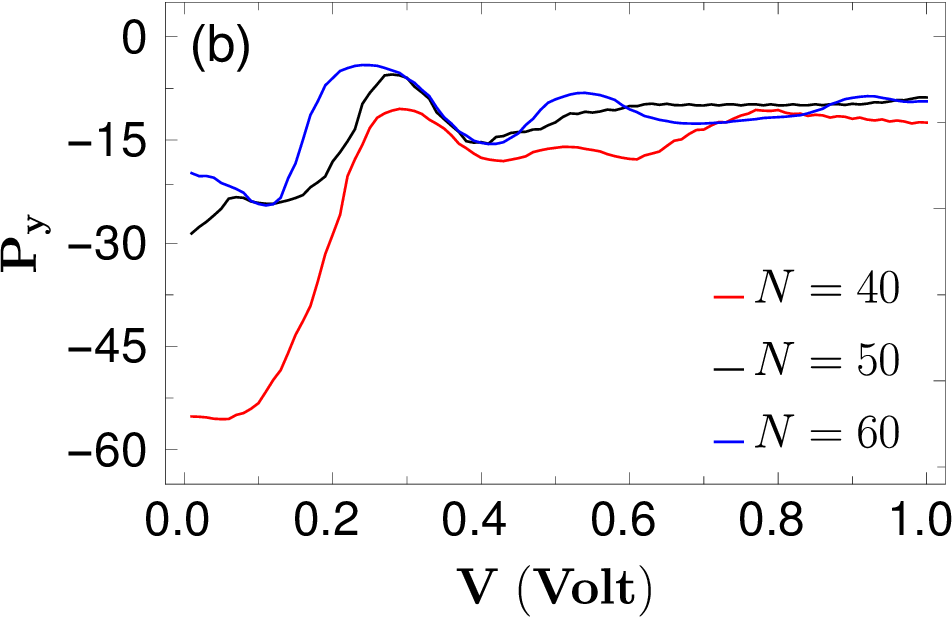}\hfill
\includegraphics[width=0.32\textwidth]{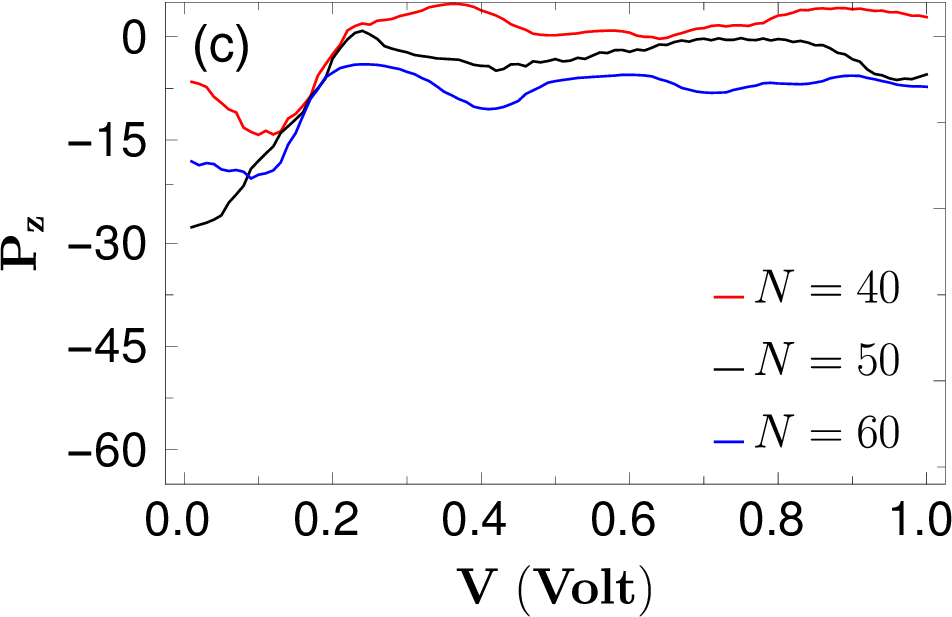}
\caption{(Color online). Bias dependence of the spin polarization coefficient $P_\alpha$ $(\alpha=x,y,z)$ for the LRH configuration for $N=40$ (denoted with red), 50 (denoted with black), and 60 (denoted with blue). Panels (a), (b), and (c) show the spin polarization coefficients $P_x$, $P_y$, and $P_z$, respectively, as a function of the applied bias voltage. Fermi energies are set at $E_F=0.4\,$eV for $N=40$, $E_F=0.5\,$eV for $N=50$, and $E_F=0.2\,$eV for $N=60$.}
\label{fig14}
\end{figure*}

To explore the finite size effect regarding the overall efficiency of the spin-polarization mechanism, in Fig.~\ref{fig14}, we illustrate the multidirectional spin polarization components ($P_x, P_y, P_z$) for $N=40$ along with two other different system sizes, namely $N=50$ and 60. Here we set the Fermi energies $E_F=0.4\,$eV for $N=40$, $E_F=0.5\,$eV for $N=50$, and $E_F=0.2\,$eV for $N=60$. As $N$ increases, the resulting spectral crowding causes a stronger overlap between the highly dense spin-up and spin-down transmission channels. Consequently, even when operating at these optimized Fermi energies, this peak overlap generally suppresses the net spin asymmetry. While the multidirectional spin polarization does not exhibit a strict, systematic scaling behavior with system size, Fig.~\ref{fig14} clearly illustrates that the overall magnitude of the spin polarization is generally reduced for larger system sizes.

Finally, alongside this coherent finite-size reduction, it is crucial to consider the physical limitations of realistic macroscopic devices. For experimental realization in extended setups, the length of the central helical system must remain smaller than or comparable to the spin coherence length of the host material. If the effective system dimension significantly exceeds this coherence length, phase-randomizing scattering events and inelastic spin-flip processes will degrade the quantum phase coherence, ultimately washing out the favorable spin-polarized response~\cite{spintronics1}.

Before we conclude, we briefly comment on the experimental feasibility and candidate materials for the proposed setup. While our tight-binding approach is phenomenological, the parameters reflect realistic experimental energy scales. Assuming a typical transition-metal hopping integral of $t \approx 1.0$-$2.5\,$eV~\cite{mehl}, our chosen exchange field $\mathpzc{h} = 0.5$ yields a physical exchange energy of roughly $0.5$-$1.25\,$eV, which is consistent with established experimental limits for exchange fields in helimagnetic alloys~\cite{turner}. For experimental realizations, the central non-collinear antiferromagnetic region could be fabricated using van der Waals helimagnets such as $\text{Ni}_{1/3}\text{NbS}_2$~\cite{kxzhang}, $3d$ transition metal compounds like $\text{FeP}$~\cite{sukhanov}, or complex pnictides (e.g.,$\text{EuNi}_{1.95}\text{As}_2$)~\cite{sangeetha2019}. The non-magnetic leads can be standard normal-metal electrodes (e.g., Au, Cu, or Pt), which offer a reliable density of states at the Fermi level and form well-characterized contact interfaces in quantum transport heterostructures.


\section{\label{sec:conclusion}Closing remarks}
In conclusion, we have investigated spin-dependent transport in a non-collinear helical antiferromagnetic system incorporating SOC within a tight-binding framework. By employing the Landauer-B\"{u}ttiker formalism, we have comprehensively analyzed the spin polarization of the transmitted electrons and examined the combined effects of the helical non-collinear magnetic texture and SOC. Distinct from prior studies that primarily attributed a finite polarization to dephasing mechanisms, our work establishes that a robust spin-splitting effect can be achieved entirely within the coherent transport regime without invoking dephasing. We have found that in the absence of SOC, the non-collinear magnetic texture alone is fully capable of generating a finite spin polarization along the $x$- and $y$-directions, whereas the $z$-component remains identically zero. Upon the inclusion of SOC, all three orthogonal components can be simultaneously obtained. Depending on the choice of system parameters, the transmitted current exhibits finite $P_x$, $P_y$, and $P_z$ components, thereby demonstrating a clear multidirectional spin filtering effect.

We have shown that both the magnitude and the sign of the spin polarization can be effectively controlled by varying the SOC strength, the Fermi energy, and the helical twisting angle. Among the two hopping regimes explored, the LRH configuration provides a substantially enhanced multidirectional spin polarization compared with the SRH case, owing to the additional transport pathways available in the LRH geometry. We have also found that the spin polarization is almost insensitive to temperature variations. These results provide deeper insights into spin transport in chiral antiferromagnetic systems and suggest that non-collinear helical antiferromagnets with SOC constitute highly promising candidates for electrically controllable multidirectional spin filters and other sophisticated spintronic applications.


\end{document}